\documentclass[aps,pra,twocolumn,showpacs,superscriptaddress]{revtex4-1}

\usepackage{amsfonts,mathrsfs,amsmath,amsthm,amssymb}
\usepackage{epsfig}
\usepackage{bm, color}

\DeclareMathAlphabet\mathbfcal{OMS}{cmsy}{b}{n}

\makeatletter
\newsavebox{\@brx}
\newcommand{\llangle}[1][]{\savebox{\@brx}{\(\m@th{#1\langle}\)}%
  \mathopen{\copy\@brx\kern-0.5\wd\@brx\usebox{\@brx}}}
\newcommand{\rrangle}[1][]{\savebox{\@brx}{\(\m@th{#1\rangle}\)}%
  \mathclose{\copy\@brx\kern-0.5\wd\@brx\usebox{\@brx}}}
\makeatother

\begin{document}  
\title {\bf 
No-go result for quantum postselection measurements 
of rank-$m$ degenerate subspace}

\author{ Le Bin Ho}
\thanks{Electronic address: binho@fris.tohoku.ac.jp}
\affiliation{Frontier Research Institute for Interdisciplinary Sciences, 
Tohoku University, Sendai 980-8578, Japan}
\affiliation{Department of Applied Physics, 
Graduate School of Engineering, 
Tohoku University, 
Sendai 980-8579, Japan}

\date{\today}

\begin{abstract}
We present a no-go result 
for postselection measurements
where the conditional expectation value
of a joint system-device observable 
under postselection is nothing else than
the conventional expectation value.
Such a no-go result relies on the rank-$m$ degenerate 
of the joint observable, where $m$ is the 
dimension of the device subspace. 
Remarkable, we show that the error and disturbance
in quantum measurements obey the no-go result,
which implies that the error-disturbance uncertainty is 
unaffected under postselection measurements.
\end{abstract}
%
%
\maketitle

{\it Introduction.}--- 
Theoretical description of quantum measurements 
with a postselection protocol 
is fundamental and of practical interest
\cite{PhysRevLett.60.1351,PhysRevA.84.052111,
PhysRevResearch.3.023243,
PhysRevLett.126.100403,PhysRevE.104.014111}. 
It is a sequential measurements 
of a generalized (POVM) measurement followed by 
another projective measurement 
\cite{PhysRevLett.60.1351}. 
The postselection process alters the statistical 
results of the measured observable 
and let to an extraordinary amplification effect: 
the expectation value obtained 
by the postselection can
go far beyond the conventional eigenvalues 
of the measured observable
\cite{PhysRevLett.60.1351,Torres2016,
PhysRevA.102.042601}. 
Beyond the fundamental interest 
\cite{RevModPhys.86.307,Qin2016,
doi:10.1098/rsta.2016.0395,
PhysRevA.100.042116,
PhysRevA.101.042117,
PhysRevA.97.012112,HO20162129,
PhysRevA.95.032135}, 
postselection measurements have attracted 
tremendous research interest in multiple fields,
including testing of quantum paradoxes
and nonlocality
\cite{PhysRevLett.102.020404,
Yokota_2009,PhysRevA.103.012228,
Calderon-Losada2020,
Xu:20,
Aharonov_2013,
Denkmayr2014,
Kim2021,Aharonov532,PhysRevA.96.052131},  
measurement uncertainty 
\cite{PhysRevA.103.022215},
weak value amplification
\cite{Hosten787,
PhysRevLett.102.173601,
PhysRevA.103.053518,
PhysRevLett.126.220801,
PhysRevA.103.053518,
ZHOU2021126655,Fang_2021,Liu_2021,
Huang2021},
quantum-enhanced metrology 
\cite{Arvidsson-Shukur2020,
PhysRevA.102.012204,Yin2021,
doi:10.1063/5.0024555,HO2019153,
PhysRevLett.118.070802,PhysRevLett.115.120401},
direct quantum state measurement
\cite{Lundeen2011,
PhysRevLett.108.070402,
PhysRevLett.121.230501,
PhysRevLett.116.040502,
PhysRevA.89.022122,
PhysRevA.97.032120,
PhysRevLett.123.150402,
Turek_2020,Ho_2020,
Tuan2021}, 
among others.

Consider a prepared state $\rho$ and a measured
observable $A$ represented by a self-adjoint operator
$\bm A$. Assume the spectral decomposition of 
$\bm A$
%
 has purely discrete spectrum as
$\bm A = \sum_kr_k\bm P_k$,
where $r_k$ is the eigenvalues, 
and projection operators 
$\bm P_k = |r_k\rangle\langle r_k|$
satisfy $\bm P_j\bm P_k = \delta_{jk}\bm P_k;
\sum_k\bm P_k = \bm I$.
Following the projection postulate
\cite{bookLR,nielsen_chuang_2010},
the expectation value gives
$\langle\bm A\rangle_\rho = \sum_k 
r_k P(r_k|\rho)$,
where $P(r_k|\rho) = {\rm Tr}[\bm P_k\rho]$ 
is the probability upon obtaining outcome $r_k$.
The state transforms to, 
following the L\"{u}ders rule \cite{bookLR,
https://doi.org/10.1002/andp.200610207}, 
$\rho' = \bm P_k\rho\bm P_k / {\rm Tr}[\bm P_k\rho]$.
After the $\bm A$-measurement, 
a subsequent projection measurement 
is carried out, i.e., using $\bm\Pi_\phi 
= |\phi\rangle\langle\phi|$, 
such that we postselect the system onto
a final state $|\phi\rangle$.
The expectation value of $\bm A$ 
now conditions on the postselected state 
$|\phi\rangle$ and reads
$_\phi\langle\bm A\rangle_\rho= \sum_k 
r_k P(r_k|\phi,\rho)$,
where $P(r_k|\phi,\rho)
= {\rm Tr}[\bm \Pi_\phi\bm P_k\rho\bm P_k]/
\sum_{k'}{\rm Tr}[\bm \Pi_\phi\bm P_{k'}\rho\bm P_{k'}]$
is the conditional probability 
following the Aharonov-Bergmann-Lebowitz 
(ABL) rule \cite{PhysRev.134.B1410}.
The conditional expectation value becomes
the weak value 
when the system weakly couples to the device.  

Even the effect of 
postsections on measurement 
results is significant,
it is not always so. 
For example, with a full degenerated spectrum,
i.e., $r_k = r\ \forall k$, or with a projective observable
$\bm A = |r\rangle\langle r|$, we have 
$\langle\bm A\rangle_\rho
=\ _\phi\langle\bm A\rangle_\rho$.
Previously, 
Vaidman et al. \cite{PhysRevA.96.032114} have claimed 
that the nature of weak values 
is the same as the eigenvalues
for an infinitesimally small interaction strength.
Besides, weak values become 
conventional expectation values
in the enlarged Hilbert space 
\cite{PhysRevA.97.012112}.

In this paper, we generalize these intuitive claims
by presenting a ``no-go" theorem, 
where the postselection 
does not affect the measurement results.
We first extend the projection postulate 
to a composite system, such as
a measured system
and its apparatus (device).
[A composite system also induces
subsystem-subsystem and 
system-environment interaction models.]
Whenever a joint system-device observable
has rank-$m$ degenerate subspace,
with $m$ is the device's dimension, 
the measured observable's results will not 
be affected by the postselection measurement. 
This is the main statement of the theorem. 
Afterward, we illustrate the no-go theorem
in the error and disturbance of quantum measurements. 
Following Ozawa's interpretation 
\cite{PhysRevA.67.042105,OZAWA2004350}, 
the error is a root-mean-square of the noise operator
formed by the device's operator after the interaction
and the system operator before the interaction,
and the disturbance in a root-mean-square of the disturbance
operator formed by the system's observables after and before the interaction.

{\it Conditional expectation values.}---
In the von Neumann mechanism \cite{neumann}, 
we consider 
a measured system $\mathcal{S}$ 
and a device $\mathcal{M}$,
initially prepared in
uncorrelated state
$|\Psi\rangle = |\psi\rangle
\otimes|\xi\rangle$.
The interaction
is given by a unitary 
$\mathbfcal U = {\rm exp}
(-it \bm H_{\mathcal{S}} \otimes
\bm H_\mathcal{M})$,
where $\bm H_{\mathcal{S}}$
and 
$\bm H_\mathcal{M}$
are Hamiltonians 
over the system's and device's complex
Hilbert spaces 
$\mathcal{H}_\mathcal{S}$ 
and $\mathcal{H}_\mathcal{M}$,
respectively. 
Given any initial joint observable 
$\mathbfcal O_0 = \bm S_0 \otimes \bm M_0$ 
in the joint $\mathcal {SM}$ 
Hilbert space,
following the Heisenberg picture,
it evolves to
$\mathbfcal O_t =
\mathbfcal U^\dagger 
\mathbfcal O_0
\mathbfcal U
$ after interaction.
Let $\mathbfcal O$
be a joint measured 
operator after interaction,
which is a function
of $\mathbfcal O_t$ 
and satiflies
$\mathbfcal{O}
\equiv f(
\mathbfcal{O}_t
)
=\sum_k \bm S_k
\otimes \bm M_k
$
 \footnote{
Using the Baker-Campbell-Hausdorff 
formula \cite{Hall2015},
we expand $
\mathbfcal U^\dagger 
\mathbfcal O_0\
\mathbfcal U
=
\mathbfcal O_0
+
it \big[\bm H_\mathcal{S}\otimes\bm H_\mathcal{M},
\mathbfcal O_0\big]
+ 
\frac{(it)^2}{2!}
\Big[\bm H_\mathcal{S}\otimes\bm H_\mathcal{M},
\big[\bm H_\mathcal{S}\otimes\bm H_\mathcal{M},
\mathbfcal O_0\big]
\Big] + 
\cdots$.
For small $t$ (or weak interaction),
it yields 
$
\mathbfcal U^\dagger 
\mathbfcal O_0\
\mathbfcal U
\approx
\mathbfcal O_0
+
it \big[\bm H_\mathcal{S}
\otimes\bm H_\mathcal{M},
\mathbfcal O_0\big]
= \sum_k
\bm S_k\otimes \bm M_k$.
}
(e.g., 
$\mathbfcal{O}
= \mathbfcal{O}_t
-\mathbfcal{O}_0$,
which pertains to the error 
and disturbance operators 
discussed later.)
To calculate the 
expectation value of 
$\mathbfcal O$, 
we start with an element
$\mathbfcal O_k \equiv 
\bm S_k\otimes \bm M_k$ 
which is an
$(n.m \times n.m)$-matrix,
where $n(m)$ is the system(device) dimension. 
Let $|u_i\rangle, |v_j\rangle$ with $i \le n$
and $j \le m$ are
eigenvectors associated 
to the eigenvalues $u_i$
and $v_j$ of 
$\bm S_k$ and $\bm M_k$, respectively,
then 
$\mathbfcal O_k$
can be expressed in the spectral representation as
$\mathbfcal O_k = \sum_{i=1}^n \sum_{j=1}^m 
r_{ij}^{(k)} \bm P_{ij}$,
where $r_{ij}^{(k)}=u_iv_j$ are eigenvalues,
and 
$\bm P_{ij}=|u_iv_j\rangle\langle u_iv_j|$
satisfies the orthonormality relations $\bm P_{ij}
\bm P_{i'j'} = \delta_{i,i'}\delta_{j,j'}\bm P_{i'j'}$
and a completeness relation $\sum \bm P_{ij} = \bm I$.
Following von Neumann, 
the probability to obtain the outcome $r_{ij}^{(k)}$ 
is given by \cite{https://doi.org/10.1002/andp.200610207}
\begin{align}\label{eq:pai}
P(r_{ij}^{(k)}|\rho) = {\rm Tr}[\bm P_{ij} \rho],
\end{align}
where we set $\rho = 
|\Psi\rangle\langle\Psi|$.
The expectation value of 
$\mathbfcal O_k$ is given by
\begin{align}\label{eq:aver}
\langle\mathbfcal O_k\rangle_\rho 
= \sum_{i,j} r_{ij}^{(k)} P(r_{ij}^{(k)}|\rho)
= \sum_{i,j} r_{ij}^{(k)} {\rm Tr}[\bm P_{ij} \rho].
\end{align}
After the projection measurement $\bm P_{ij}$,
the joint state 
transforms to a conditional 
(not normalized)
$\rho'_{ij} = \bm P_{ij}\rho\bm P_{ij}$.
We then postselect system $\mathcal S$ 
onto  a final state
$|\phi\rangle$, represented by 
a projection operator $\bm \Pi_\phi 
= |\phi\rangle\langle\phi| \otimes \bm I$.
The joint probability to obtain $r_{ij}^{(k)}$ 
and postselection is
\begin{align}\label{eq:jp}
P(r_{ij}^{(k)},\phi|\rho) = {\rm Tr}[\bm \Pi_\phi
\rho'_{ij}]
={\rm Tr}[\bm \Pi_\phi
\bm P_{ij}\rho\bm P_{ij}].
\end{align}
Using the Bayesian theorem,
the conditional probability to obtain
$r_{ij}^{(k)}$ for 
given pre and postselected states is
\begin{align}\label{eq:conp}
\notag P(r_{ij}^{(k)}|\phi,\rho) &= 
\dfrac{P(r_{ij}^{(k)},\phi|\rho)}
{\sum_{i',j'} P(r_{i'j'}^{(k)},\phi|\rho)} \\
&=\dfrac{{\rm Tr}[\bm \Pi_\phi
\bm P_{ij}\rho\bm P_{ij}]}
{\sum_{i',j'} {\rm Tr}[\bm \Pi_\phi
\bm P_{i'j'}\rho\bm P_{i'j'}]}.
\end{align}
Then, the conditional expectation value
of 
$\mathbfcal O_k$ yields 
\begin{align}\label{eq:aveA}
_\phi\langle\mathbfcal O_k\rangle_\rho
= \sum_{i,j} r_{ij}^{(k)} P(r_{ij}|\phi,\rho)
= \dfrac{\sum_{i,j} r_{ij}^{(k)}{\rm Tr}[\bm \Pi_\phi
\bm P_{ij}\rho\bm P_{ij}]}
{\sum_{i',j'} {\rm Tr}[\bm \Pi_\phi
\bm P_{i'j'}\rho\bm P_{i'j'}]}.
\end{align}

%
We present the following theorem:\\
{\it Theorem (no-go postselection theorem).}--- 
For any given 
joint state $\rho = |\Psi\rangle\langle\Psi|$ 
and post-selected
state $|\phi\rangle$,
the following rank-$m$ 
degenerate 
for every joint operator $\mathbfcal{O}_k$
\begin{align}\label{eq:de}
r_{ij}^{(k)} = r_{i'j}^{(k)} \equiv \widetilde{r}_j^{(k)}, 
\forall\ 1 \le i,i' \le n, 
\text{ and } 1 \le j \le m,
\end{align}
must lead to a 
``no-go" for postselection
measurement 
\begin{align}\label{eq:nogo}
_\phi\langle \mathbfcal O_k\rangle_\rho
=\langle\mathbfcal O_k\rangle_\rho
\text{ and }
_\phi\langle \mathbfcal O\rangle_\rho
=\langle\mathbfcal O\rangle_\rho,
\end{align}
where $\langle\mathbfcal O\rangle
=\sum_k \langle\mathbfcal O_k\rangle$.

{\it Proof of Theorem.}---
Let $\{|e_i\rangle\otimes|g_j\rangle\}$
be the canonical basis of the joint space,
in which the joint state $|\Psi\rangle$
can be expressed as
\begin{align}
|\Psi\rangle=
	\begin{pmatrix}
	\psi_1 \\ \psi_2 \\ \vdots \\ \psi_n
	\end{pmatrix}
\otimes
	\begin{pmatrix}
	\xi_1 \\ \xi_2 \\ \vdots \\ \xi_m
	\end{pmatrix}
=\sum_{i =1}^n\sum_{j=1}^m
\psi_i\xi_j|e_ig_j\rangle,
\end{align}
where $\psi_i = \langle e_i|\psi\rangle$
and $\xi_j = \langle g_j|\xi\rangle$.
Let the eigenvalues 
$|r_{ij}^{(k)}\rangle \equiv |u_iv_j\rangle = 
\sum_{i',j'} a_{i'j',ij}
|e_{i'}g_{j'}\rangle$
is the eigenbasis,
and $\bm T = (|r_{11}^{(k)}\rangle, 
\cdots, |r_{nm}^{(k)}\rangle)$
is the transformation matrix
that formed by the ket vector of all eigenvalues. 
In the eigenbases, the joint state is expressed as
$|\Psi\rangle = 
\sum_{i,j} \psi'_i\xi'_j |u_iv_j\rangle$,
where $\psi'_i = \langle u_i|\psi\rangle$
and $\xi'_j = \langle v_j|\xi\rangle$,
whose obey $\sum_i |\psi'_i|^2 = 1$
and $\sum_j |\xi'_j|^2 = 1$. We then obtain
$\langle r_{ij}^{(k)}|\Psi\rangle
=\psi_i'\xi_j'$.
 
In the postseleted projection operator 
$\bm \Pi_\phi 
= |\phi\rangle\langle\phi| \otimes \bm I$,
let $|\phi\rangle = 
\sum_i\phi_i |e_i\rangle$, 
where $\phi_i = \langle e_i|\phi\rangle$
a complex amplitude, 
then we obtain $(n.m \times n.m)$-matrix:
\begin{align}\label{eq:pi}
\bm \Pi_\phi = 
\begin{pmatrix}
\newcommand{\lr}[1]{\multicolumn{1}{|c}{#1}}
\newcommand{\rr}[1]{\multicolumn{1}{c|}{#1}}
\;
\begin{array}{@{}*{10}{c}@{}}
\cline{1-3}
\lr{|\phi_1|^2} & & \rr{} \\ 
\lr{} &\ddots & \rr{} \\
\lr{} & & \rr{|\phi_1|^2}\\
\cline{1-3}
&&& \ddots \\
\cline{6-8}
&&&&& \lr{|\phi_n|^2} & & \rr{} \\
&&&&& \lr{} &\ddots & \rr{} \\
&&&&& \lr{} & & \rr{|\phi_n|^2} \\
\cline{6-8}
\end{array}
\end{pmatrix},
\end{align}
where the off-diagonal elements
are omitted as they do not 
include in the canonical basis,
see more details in App.~\ref{appA}.
Here, each box is an $(m\times m)$-matrix,
with totally $n$ boxes. In the eigenbasis, 
the postselected state is expressed as
$\bm \Pi'_\phi = \bm T^\dagger 
\bm \Pi_\phi\bm T$.
An extra requirement (for the transformation matrix)
that the diagonal elements
of $\bm \Pi'_\phi$ admit the rank-$m$ degenerated 
subspace similar as $\bm \Pi_\phi$, 
i.e., ${\rm diag}(\bm \Pi'_\phi) = 
(|\phi'_1|^2, \cdots, |\phi'_1|^2,\cdots, 
|\phi'_n|^2, \cdots, |\phi'_n|^2)$.
We emphasize that this requirement 
always satisfies
when the eigenbasis is the canonical basis
as given in Eq.~\eqref{eq:pi}.

Now, from Eq.~\eqref{eq:aveA}, 
we have the denominator
\begin{align}\label{eq:msf}
\sum_{i,j=1}^{n,m} {\rm Tr}
[\bm \Pi_\phi \bm P_{ij} \rho \bm P_{ij}]
= \sum_{i=1}^n |\phi'_i|^2 \psi'_i|^2,
\end{align}
and the numerator
\begin{align}\label{eq:ts}
 \sum_{i,j} r_{ij}^{(k)} {\rm Tr}
[\bm \Pi_\phi \bm P_{ij} \rho \bm P_{ij}]
= \sum_{j=1}^m {\widetilde r}_j^{(k)}|\xi'_j|^2 \cdot 
\sum_{i=1}^n |\phi'_i|^2 \psi'_i|^2,
\end{align}
where 
we applied condition~\eqref{eq:de}.
Equation \eqref{eq:aveA} is recast as
\begin{align}\label{eq:aveAre}
_\phi\langle\mathbfcal O_k\rangle_\rho
= \sum_{j=1}^m {\widetilde r}_j^{(k)}|\xi'_j|^2.
\end{align}
Similarly, we have 
$\langle\mathbfcal O_k\rangle_\rho
= \sum_{j=1}^m {\widetilde r}_j^{(k)}|\xi'_j|^2$,
then $_\phi\langle\mathbfcal O_k\rangle_\rho
=\langle\mathbfcal O_k\rangle_\rho$
(See detailed proof in App.~\ref{appA}
.)
The proof for second term in \eqref{eq:nogo}
is straightforward since
all $\mathbfcal{O}_k$ satisfy 
condition \eqref{eq:de}
$\;\; \Box$

{\it Corollary 1.}---
For any eigenbasis $\{|r_{ij}\rangle\}$ 
is the canonical basis, 
the no-go theorem states
\begin{align}\label{eq:nogoc}
_\phi\langle\mathbfcal O\rangle_\rho
=\langle\mathbfcal O\rangle_\rho
= \sum_{j=1}^m {\widetilde r}_j|\xi_j|^2.
\end{align}
The proof for this {\it Corollary} is 
the same as above.


{\it Remark.}---
Different from conditional expectation values 
we are considering here, 
the weak value of an observable $\bm A$
in system $\mathcal S$ generally depends 
on the postselected state,
as it is $\langle\bm A\rangle_{\rm w}
= {\langle\phi|\bm A|\psi\rangle}/
{\langle\phi|\psi\rangle}$,
except for some certain conditions wherein
it can be an expectation value
\cite{PhysRevA.97.012112} 
or an eigenvalue \cite{PhysRevA.96.032114}.
However, this is not a consequential result
from our theorem here. 
Instead, a consequence from 
the no-go theorem can state as follows:

{\it Corollary 2.}---
Weak values can reduce to eigenvalues 
if the measured observable $\bm A$
has  full-rank degenerate subspace. 
To proof this {\it Corollary}, let's say
$\bm A = \sum_k a_k |k\rangle\langle k|$
with $a_k = a$ for all $k = 1, \cdots, n$,
then, $\langle\bm A\rangle_{\rm w} = 
\langle \bm A\rangle = a$.

{\it Observation.}---
Any joint observable, 
i.e., $\mathbfcal O = 
\bm S\otimes\bm M$
with $\bm S = \bm I$ 
always satisfies 
the no-go theorem. 
The proof for this 
{\it Observation} 
is given directly by noting 
that the eigenvalues of 
$\bm I$ are all one.
Thus eigenvalues
of  $\mathbfcal O $ 
satisfies \eqref{eq:de},
and thus satisfies the theorem.

{\it No-go theorem in the error and disturbance.}---
Error and disturbance 
are essential quantities for determining 
measurements' uncertainties
\cite{
PhysRevA.67.042105,
OZAWA2004350,
Ozawa_2005}.
We consider 
the error of an $\bm{A}$-measurement 
in system $\mathcal S$
through an $\bm{M}$-measurement 
in device $\mathcal M$
and the disturbance of a 
$\bm{B}$-measurement 
in system $\mathcal S$.
In the joint $\mathcal {SM}$ system, 
we denote 
$\mathbfcal A_0 
= \bm{A}\otimes\bm{I}, \text{ and } 
\mathbfcal B_0 
= \bm{B}\otimes\bm{I}$,
where $\bm{A}$ and $\bm{B}$ 
are the observables to be measured 
in system $\mathcal S$. We also define 
a device observable $\bm{M}$ 
in the device space, 
such that it becomes $\mathbfcal M_0 = 
\bm{I}\otimes\bm{M}$ in the 
joint $\mathcal {SM}$ space
\cite{PhysRevA.67.042105,
OZAWA2004350,
Ozawa_2005}.
The interaction is switched on during 
a short time $t$, where the
joint $\mathcal {SM}$ system 
evolves under the unitary transformation 
$\mathbfcal U$. These operators transform to 
$\mathbfcal M_t = 
\mathbfcal U^\dagger
\mathbfcal M_0\mathbfcal U$
and $\mathbfcal B_t = \mathbfcal U^\dagger
\mathbfcal B_0\mathbfcal U$.
According to Ozawa~\cite{
PhysRevA.67.042105,
OZAWA2004350,
Ozawa_2005},   
the error and disturbance operators 
are defined by 
$\mathbfcal N_{\bm A} = 
\mathbfcal M_t - \mathbfcal A_0,$
and  $\mathbfcal D_{\bm B} =
\mathbfcal B_t - \mathbfcal B_0$,
respectively. 
Then, the mean square error and  
the disturbance are given by
\begin{align}\label{error_p}
\epsilon^2_{\bm A} = \langle 
\mathbfcal N_{\bm A}^2\rangle_\Psi, 
\text{ and } 
\eta_{\bm B}^2
= \langle \mathbfcal D^2_{\bm B}\rangle_\Psi,
\end{align}
where the bra-ket symbol 
$\langle ... \rangle_\Psi$ means 
$\langle\Psi|...|\Psi\rangle$ 
throughout this paper.

In the following, we illustrate 
that such the error and disturbance 
satisfy condition~\eqref{eq:de} in
the {\it Theorem}, and thus
it makes no sence for postselection
measurements of the error
and disturbance. In other words,
the error and disturbance will 
not be affected under the postselection.

Concretely, let us consider a CNOT-type measurement,
where both 
system $\mathcal S$ and device $\mathcal M$
are qubits, initially prepared in 
$|\psi\rangle = |i^+\rangle$
and $|\xi\rangle = \sqrt{\frac{1+s}{2}}|0\rangle
+\sqrt{\frac{1-s}{2}} |1\rangle$, 
where $|0\rangle$ and $|1\rangle$ 
are two eigenstates
of Pauli matrix $\bm Z$, and 
 $|i^+\rangle = \frac{1}{\sqrt{2}} (|0\rangle+i|1\rangle)$; 
$s$ is the measurement strength ranging from 
0 (weak measurement) to 1 (strong measurement).
The measured observables $\bm A$ and $\bm B$
in system $\mathcal S$ are chosen t
o be Pauli matrices 
$\bm Z$ and $\bm X$, respectively,
and the device observable $\bm M$
is also $\bm Z$. 
The interaction is CNOT-gate, i.e.,
$\mathbfcal U = |0\rangle\langle0|\otimes \bm I
+ |1\rangle\langle1|\otimes \bm X$.
The square error 
and disturbance operators give
\begin{align}\label{eq:sesd}
\mathbfcal N^2_{\bm Z} 
= 4\bm I \otimes |1\rangle\langle1|,
\text{ and } 
\mathbfcal D^2_{\bm X} 
= 2\bm I \otimes (\bm I -\bm X).
\end{align}

Fortunately, CNOT is a typical interaction,
which leads to simplify measured operators
(square error and disturbance).
The eigenvalues of the square error
are (0, 4, 0, 4), and the same for
the square disturbance, which all satisfy 
the rank-2 degenerate.
As a result, 
$_\phi\langle\mathbfcal N^2_{\bm Z}\rangle_\Psi 
= \langle\mathbfcal N^2_{\bm Z}\rangle_\Psi
= 2(1-s),$ 
and 
$_\phi\langle\mathbfcal D^2_{\bm X}\rangle_\Psi 
= \langle\mathbfcal D^2_{\bm X}\rangle_\Psi
= 2(1-\sqrt{1-s^2}),$ 
for any postselected state $|\phi\rangle = \cos\theta|0\rangle
+ e^{-i\varphi} \sin\theta |1\rangle$.
These results imply that postselection measurements 
affect neither the error nor the disturbance.
(See detailed calculation in App.~\ref{appB}.)

The observations in the example
can be explained as follows. 
The error is determined via the device 
after the first measurement, and thus, 
it is not affected by postselection measurements
as long as the rank-$m$ degenerate 
in the device holds.
For the same reason, the disturbance is affected 
by the backaction caused by the device 
while unpolluted 
with postselection measurements.

{\it Conclusion.}--- 
We introduced a no-go theorem
for postselection measurements,
where the obtained conditional expectation value
is not affected by postselection measurements
and is equal to a conventional expectation value.
This can happen when the joint observable has 
a rank-$m$ degenerate subspace, where $m$ is the dimension 
of the device space. 
As a consequence, the error and disturbance in 
quantum measurements  immune with
postselection measurements.
  
{\it Acknowledgments.}---
We thanks H.C. Nguyen for the fruitful discussion. 

\appendix
\setcounter{equation}{0}
\renewcommand{\theequation}{A.\arabic{equation}}
\section{Detailed proof for the no-go theorem}\label{appA}
In this proof, we omit the indicator $k$ for short.
We first derive ${\rm Tr}
[\bm \Pi_\phi \bm P_{ij} \rho \bm P_{ij}]$
in Eq.~\eqref{eq:msf}
in the main text. We have
\begin{align}\label{app:eq:tr}
\notag {\rm Tr}[\bm \Pi_\phi \bm P_{ij} \rho \bm P_{ij}]
& = \langle \Psi|\bm P_{ij}\bm\Pi_\phi\bm P_{ij}|\Psi\rangle\\
& = \langle \Psi|r_{ij}\rangle
\langle r_{ij}|
\bm\Pi_\phi|r_{ij}\rangle
\langle r_{ij}
|\Psi\rangle,
\end{align}
where we used  
$\bm P_{ij} = |r_{ij}\rangle
\langle r_{ij}|$.
Concretely, we derive 
\begin{align}\label{app:eq:pst}
\langle r_{ij} |\Psi\rangle 
= \langle u_i v_j
|\psi\xi\rangle
 = \psi_i'\xi_j',
\end{align}
where we set 
$\psi_i' = \langle u_i|\psi\rangle, 
\xi_j' = \langle v_j|\xi\rangle$, and 
\begin{align}\label{app:eq:phi'}
\langle r_{ij}|\bm\Pi_\phi|r_{ij}\rangle
= \langle e_ig_j| \bm T^\dagger\bm \Pi_\phi
\bm T |e_ig_j\rangle
 = (\bm \Pi'_\phi )_{ij\times ij},
\end{align}
where we applied 
the bases transformation rule
$|r_{ij}\rangle = \bm T|e_ig_j\rangle$,
and set $\bm \Pi'_\phi = \bm T^\dagger\bm \Pi_\phi
\bm T$.
We also consider the case
${\rm diag}(\bm \Pi'_\phi) = 
(|\phi'_1|^2, \cdots, |\phi'_1|^2,\cdots, 
|\phi'_n|^2, \cdots, |\phi'_n|^2)$.
Then, substituting Eqs.~(\ref{app:eq:pst}, \ref{app:eq:phi'})
into Eq.~\eqref{app:eq:tr}, we obtain Eq.~\eqref{eq:msf}

\begin{align}\label{app:eq:msf}
\notag \sum_{i,j=1}^{n,m} {\rm Tr}
[\bm \Pi_\phi \bm P_{ij} \rho \bm P_{ij}]
& = \sum_{i,j=1}^{n,m} 
\langle \Psi|r_{ij}\rangle
\langle r_{ij}|
\bm\Pi_\phi|r_{ij}\rangle
\langle r_{ij}
|\Psi\rangle\\
\notag & = \sum_{i,j=1}^{n,m} 
|\psi'_i|^2\ |\xi_j'|^2\
|\phi_i'|^2\\
& = \sum_{i=1}^n |\psi'_i|^2\ |\phi'_i|^2.
\end{align}

Next, we derive Eq.~\eqref{eq:ts}
in the main text
\begin{align}\label{app:eq:ts}
\notag \sum_{i,j} r_{ij} {\rm Tr}
[\bm \Pi_\phi \bm P_{ij} \rho \bm P_{ij}]
& = \sum_{i,j=1}^{n,m}
r_{ij}\ |\psi'_i|^2 \ |\xi'_j|^2 \
|\phi'_i|^2 \\
& = \sum_{i=1}^n |\psi'_i|^2\ |\phi'_i|^2
\cdot \sum_{j=1}^m {\widetilde r}_j\ |\xi'_j|^2.
\end{align}
Finally, the conditional expectation value
of the observable $\mathbfcal O$ is given by
\begin{align}\label{appeq:aveA}
_\phi\langle\mathbfcal O\rangle_\rho
= \dfrac{\sum_{i,j} r_{ij}{\rm Tr}[\bm \Pi_\phi
\bm P_{ij}\rho\bm P_{ij}]}
{\sum_{i',j'} {\rm Tr}[\bm \Pi_\phi
\bm P_{i'j'}\rho\bm P_{i'j'}]} 
= \sum_{j=1}^m {\widetilde r}_j\ |\xi'_j|^2.
\end{align}
We compare the conditional expectation value
with the conventional expectation value in
Eq.~\eqref{eq:aver} in the main text:
\begin{align}\label{eq:aver}
\notag \langle\mathbfcal O\rangle_\rho 
& = \sum_{i,j} r_{ij} {\rm Tr}[\bm P_{ij} \rho]\\
\notag & = \sum_{i,j} \widetilde r_j \langle\Psi|r_{ij}
\rangle\langle r_{ij}|\Psi\rangle\\
\notag & = \sum_{i,j} \widetilde r_j \
|\psi'_i|^2\ |\xi'_j|^2\\
& = \sum_j \widetilde r_j\ |\xi'_j|^2,
\end{align}
and obtain $_\phi\langle\mathbfcal O\rangle_\rho
=\langle\mathbfcal O\rangle_\rho$,
which completes the proof. 

\setcounter{equation}{0}
\renewcommand{\theequation}{B.\arabic{equation}}
\section{Error and disturbance in 
CNOT-type measurement}\label{appB}
In this section, we give a detailed calculation of 
square error 
and square disturbance. 
First, we explicitly derive the initial joint state as
\begin{align}\label{eq:app_Psi}
|\Psi\rangle = |\psi\rangle\otimes |\xi\rangle
=\dfrac{1}{2}
\begin{pmatrix}
\sqrt{1+s}\\
\sqrt{1-s}\\
i\sqrt{1+s}\\
i\sqrt{1-s}
\end{pmatrix},
\end{align}
here, $\psi_1 = \psi_2 = 1/\sqrt{2}$,
and $\xi_1 = \sqrt{\frac{1+s}{2}}, 
\xi_2 = \sqrt{\frac{1-s}{2}}$.
The square error operator 
in Eq.~\eqref{eq:sesd} is decomposed 
into its eigenvalue and eigenstate as
\begin{align}\label{eq:appsesd}
\notag \mathbfcal N^2_{\bm Z} 
&= 4\bm I \otimes |1\rangle\langle1|\\
\notag &= 0|00\rangle\langle 00| 
+ 4|01\rangle\langle 01| 
+ 0|10\rangle\langle 10| 
+ 4|11\rangle\langle 11|\\
&\equiv \sum_{i,j} r_{ij}\bm P_{ij}.
\end{align}
Here, obviously, we have 
$r_{11} = r_{21} = 0\ (\equiv \widetilde r_1)$
and $r_{12} = r_{22} = 4 \ (\equiv \widetilde r_2)$, 
which satisfy condition \eqref{eq:de}.
The projectors are 
$\bm P_{11} = {\rm diag}(1, 0, 0, 0); 
\bm P_{12} = {\rm diag}(0, 1, 0, 0);
\bm P_{21} = {\rm diag}(0, 0, 1, 0);$ 
and $\bm P_{22} = {\rm diag}(0, 0, 0, 1)$.
The post-selected state explicitly reads
\begin{align}\label{eq:app_pi}
\bm \Pi_\phi = 
   \begin{pmatrix}
   \cos^2\theta & & & \\
   & \cos^2\theta & & \\
   & & \sin^2\theta &\\
   & & & \sin^2\theta
   \end{pmatrix},
\end{align}
while we ignore the off-diagonal terms,
hence $\phi_1 =\cos\theta$, 
$\phi_2 = e^{-i\varphi}\sin\theta$.
We emphasize that in this case,
the eigenbasis is also the canonical basis,
such that $\bm T = \bm I$.
Now, we calculate \eqref{eq:msf}
\begin{align}\label{eq:appmsf}
\sum_{i,j=1}^{2,2} {\rm Tr}
[\bm \Pi_\phi \bm P_{ij} \rho \bm P_{ij}]
= \sum_{i=1}^2 |\phi_i|^2\ |\psi_i|^2
= \dfrac{1}{2}.
\end{align}
Noting that direct calculating the L.H.S 
gives us exactly the same result.
Next, we evaluate \eqref{eq:ts}
\begin{align}\label{eq:app_ts}
\notag \sum_{i,j} r_{ij} {\rm Tr}
[\bm \Pi_\phi \bm P_{ij} \rho \bm P_{ij}]
& = \sum_{j=1}^2 {\widetilde r}_j\ |\xi_j|^2 \cdot 
\sum_{i=1}^2 |\phi_i|^2\ |\psi_i|^2 \\
& = 1-s.
\end{align}
Again, direct calculating the L.H.S 
gives us exactly the same result.
Now, Eq.~\eqref{eq:aveA} explicitly reads 
\begin{align}\label{eq:app_aveA}
_\phi\langle\mathbfcal N^2_{\bm Z} \rangle_\rho
= 2(1-s),
\end{align}
which is the square error under 
the pre- and postselection measurement scheme.
Furthermore, in the usual case without postselection,
the square error also reads 
$\langle\mathbfcal 
N^2_{\bm Z} \rangle_\rho
= 2(1-s)$, which implies no-go for postselection
as shown in Eq.~\eqref{eq:nogoc} of {\it Corollary 1}.

Now, we evaluate the square disturbance.
The square disturbance operator 
states
in Eq.~\eqref{eq:sesd} is decomposed 
into its eigenvalue and eigenstate as
\begin{align}\label{eq:app_dis}
\notag \mathbfcal D^2_{\bm X} 
&= 2\bm I \otimes (\bm I - \bm X)\\
\notag &= 0|0+\rangle\langle 0\!+\!| 
+ 4|0-\rangle\langle 0\!-\!| \\
&\hspace{2cm}+ 0|1
+\rangle\langle 1\!+\!| 
+ 4|1-\rangle\langle 1\!-\!|.
\end{align}
Here, $|0+\rangle = |0\rangle\otimes |+\rangle$
and likewise for the others, and $|\pm\rangle = 
(|0\rangle\pm|1\rangle)/\sqrt{2}.$
Similar as the square-error case, 
we have 
$r_{11} = r_{21} = 0\ (\equiv \widetilde r_1)$
and $r_{12} = r_{22} = 4 \ (\equiv \widetilde r_2)$, 
which satisfy condition \eqref{eq:de}.
The projectors in this case are
$\bm P_{11} = |0+\rangle\langle0\!+\!|; 
\bm P_{12} = |0-\rangle\langle0\!-\!|;
\bm P_{21} = |1+\rangle\langle1\!+\!|;$ 
and $\bm P_{22} = |1-\rangle\langle1\!-\!|$.

We next introduce the transformation matrix
\begin{align}\label{eq:app_T}
\notag \bm T &=  
	\begin{pmatrix}
	|0+\rangle, & |0-\rangle,&
|1+\rangle, & |1-\rangle
	\end{pmatrix} \\
&=\dfrac{1}{\sqrt{2}}
	\begin{pmatrix}
	1 & -1 & 0 & 0\\
	1 & 1 & 0 & 0\\
	0 & 0 & 1 & -1\\
	0 & 0 & 1 & 1
	\end{pmatrix}.
\end{align}
The postselected state 
in the eigenbasis reads
\begin{align}\label{eq:app_pi}
\bm \Pi'_\phi = \bm T^\dagger 
\bm \Pi_\phi\bm T
= {\rm diag}\bigl(\cos^2\theta, \cos^2\theta, 
\sin^2\theta, \sin^2\theta\bigr),
\end{align}
where we have ignored the off-diagonal terms
as it will not affect under the eigenbasis.
Now, we calculate \eqref{eq:msf}
by inserting $\bm T\bm T^\dagger $
between operators inside the trade:
\begin{align}\label{eq:appmsfD}
\sum_{i,j=1}^{2,2} {\rm Tr}
[\bm \Pi'_\phi \bm P'_{ij} \rho' \bm P'_{ij}]
= \sum_{i=1}^2 |\phi'_i|^2\ |\psi'_i|^2
= \dfrac{1}{2}.
\end{align}
Noting that direct calculating the L.H.S 
gives us exactly the same result.
Next, we evaluate \eqref{eq:ts}
\begin{align}\label{eq:app_ts}
\notag \sum_{i,j} r_{ij} {\rm Tr}
[\bm \Pi_\phi \bm P_{ij} \rho \bm P_{ij}]
& = \sum_{j=1}^2 {\widetilde r}_j\ |\xi'_j|^2 \cdot 
\sum_{i=1}^2 |\phi'_i|^2\ |\psi'_i|^2 \\
& = 1-\sqrt{1-s^2}.
\end{align}
Now, Eq.~\eqref{eq:aveA} explicitly reads 
\begin{align}\label{eq:app_aveAD}
_\phi\langle\mathbfcal D^2_{\bm X} \rangle_\rho
= 2(1-\sqrt{1-s^2}),
\end{align}
which is the square disturbance under 
the pre- and postselection measurement scheme.
Furthermore, in the usual case without postselection,
the square error also reads 
$\langle\mathbfcal 
D^2_{\bm X} \rangle_\rho
= 2(1-\sqrt{1-s^2})$, which again  
implies no-go for postselection.

\bibliography{refs}

\begin{thebibliography}{62}%
\makeatletter
\providecommand \@ifxundefined [1]{%
 \@ifx{#1\undefined}
}%
\providecommand \@ifnum [1]{%
 \ifnum #1\expandafter \@firstoftwo
 \else \expandafter \@secondoftwo
 \fi
}%
\providecommand \@ifx [1]{%
 \ifx #1\expandafter \@firstoftwo
 \else \expandafter \@secondoftwo
 \fi
}%
\providecommand \natexlab [1]{#1}%
\providecommand \enquote  [1]{``#1''}%
\providecommand \bibnamefont  [1]{#1}%
\providecommand \bibfnamefont [1]{#1}%
\providecommand \citenamefont [1]{#1}%
\providecommand \href@noop [0]{\@secondoftwo}%
\providecommand \href [0]{\begingroup \@sanitize@url \@href}%
\providecommand \@href[1]{\@@startlink{#1}\@@href}%
\providecommand \@@href[1]{\endgroup#1\@@endlink}%
\providecommand \@sanitize@url [0]{\catcode `\\12\catcode `\$12\catcode
  `\&12\catcode `\#12\catcode `\^12\catcode `\_12\catcode `\%12\relax}%
\providecommand \@@startlink[1]{}%
\providecommand \@@endlink[0]{}%
\providecommand \url  [0]{\begingroup\@sanitize@url \@url }%
\providecommand \@url [1]{\endgroup\@href {#1}{\urlprefix }}%
\providecommand \urlprefix  [0]{URL }%
\providecommand \Eprint [0]{\href }%
\providecommand \doibase [0]{http://dx.doi.org/}%
\providecommand \selectlanguage [0]{\@gobble}%
\providecommand \bibinfo  [0]{\@secondoftwo}%
\providecommand \bibfield  [0]{\@secondoftwo}%
\providecommand \translation [1]{[#1]}%
\providecommand \BibitemOpen [0]{}%
\providecommand \bibitemStop [0]{}%
\providecommand \bibitemNoStop [0]{.\EOS\space}%
\providecommand \EOS [0]{\spacefactor3000\relax}%
\providecommand \BibitemShut  [1]{\csname bibitem#1\endcsname}%
\let\auto@bib@innerbib\@empty
\bibitem [{\citenamefont {Aharonov}\ \emph {et~al.}(1988)\citenamefont
  {Aharonov}, \citenamefont {Albert},\ and\ \citenamefont
  {Vaidman}}]{PhysRevLett.60.1351}%
  \BibitemOpen
  \bibfield  {author} {\bibinfo {author} {\bibfnamefont {Y.}~\bibnamefont
  {Aharonov}}, \bibinfo {author} {\bibfnamefont {D.~Z.}\ \bibnamefont
  {Albert}}, \ and\ \bibinfo {author} {\bibfnamefont {L.}~\bibnamefont
  {Vaidman}},\ }\href {\doibase 10.1103/PhysRevLett.60.1351} {\bibfield
  {journal} {\bibinfo  {journal} {Phys. Rev. Lett.}\ }\textbf {\bibinfo
  {volume} {60}},\ \bibinfo {pages} {1351} (\bibinfo {year}
  {1988})}\BibitemShut {NoStop}%
\bibitem [{\citenamefont {Zhu}\ \emph {et~al.}(2011)\citenamefont {Zhu},
  \citenamefont {Zhang}, \citenamefont {Pang}, \citenamefont {Qiao},
  \citenamefont {Liu},\ and\ \citenamefont {Wu}}]{PhysRevA.84.052111}%
  \BibitemOpen
  \bibfield  {author} {\bibinfo {author} {\bibfnamefont {X.}~\bibnamefont
  {Zhu}}, \bibinfo {author} {\bibfnamefont {Y.}~\bibnamefont {Zhang}}, \bibinfo
  {author} {\bibfnamefont {S.}~\bibnamefont {Pang}}, \bibinfo {author}
  {\bibfnamefont {C.}~\bibnamefont {Qiao}}, \bibinfo {author} {\bibfnamefont
  {Q.}~\bibnamefont {Liu}}, \ and\ \bibinfo {author} {\bibfnamefont
  {S.}~\bibnamefont {Wu}},\ }\href {\doibase 10.1103/PhysRevA.84.052111}
  {\bibfield  {journal} {\bibinfo  {journal} {Phys. Rev. A}\ }\textbf {\bibinfo
  {volume} {84}},\ \bibinfo {pages} {052111} (\bibinfo {year}
  {2011})}\BibitemShut {NoStop}%
\bibitem [{\citenamefont {Wagner}\ \emph {et~al.}(2021)\citenamefont {Wagner},
  \citenamefont {Kersten}, \citenamefont {Danner}, \citenamefont {Lemmel},
  \citenamefont {Pan},\ and\ \citenamefont
  {Sponar}}]{PhysRevResearch.3.023243}%
  \BibitemOpen
  \bibfield  {author} {\bibinfo {author} {\bibfnamefont {R.}~\bibnamefont
  {Wagner}}, \bibinfo {author} {\bibfnamefont {W.}~\bibnamefont {Kersten}},
  \bibinfo {author} {\bibfnamefont {A.}~\bibnamefont {Danner}}, \bibinfo
  {author} {\bibfnamefont {H.}~\bibnamefont {Lemmel}}, \bibinfo {author}
  {\bibfnamefont {A.~K.}\ \bibnamefont {Pan}}, \ and\ \bibinfo {author}
  {\bibfnamefont {S.}~\bibnamefont {Sponar}},\ }\href {\doibase
  10.1103/PhysRevResearch.3.023243} {\bibfield  {journal} {\bibinfo  {journal}
  {Phys. Rev. Research}\ }\textbf {\bibinfo {volume} {3}},\ \bibinfo {pages}
  {023243} (\bibinfo {year} {2021})}\BibitemShut {NoStop}%
\bibitem [{\citenamefont {Monroe}\ \emph {et~al.}(2021)\citenamefont {Monroe},
  \citenamefont {Yunger~Halpern}, \citenamefont {Lee},\ and\ \citenamefont
  {Murch}}]{PhysRevLett.126.100403}%
  \BibitemOpen
  \bibfield  {author} {\bibinfo {author} {\bibfnamefont {J.~T.}\ \bibnamefont
  {Monroe}}, \bibinfo {author} {\bibfnamefont {N.}~\bibnamefont
  {Yunger~Halpern}}, \bibinfo {author} {\bibfnamefont {T.}~\bibnamefont {Lee}},
  \ and\ \bibinfo {author} {\bibfnamefont {K.~W.}\ \bibnamefont {Murch}},\
  }\href {\doibase 10.1103/PhysRevLett.126.100403} {\bibfield  {journal}
  {\bibinfo  {journal} {Phys. Rev. Lett.}\ }\textbf {\bibinfo {volume} {126}},\
  \bibinfo {pages} {100403} (\bibinfo {year} {2021})}\BibitemShut {NoStop}%
\bibitem [{\citenamefont {Purves}\ and\ \citenamefont
  {Short}(2021)}]{PhysRevE.104.014111}%
  \BibitemOpen
  \bibfield  {author} {\bibinfo {author} {\bibfnamefont {T.}~\bibnamefont
  {Purves}}\ and\ \bibinfo {author} {\bibfnamefont {A.~J.}\ \bibnamefont
  {Short}},\ }\href {\doibase 10.1103/PhysRevE.104.014111} {\bibfield
  {journal} {\bibinfo  {journal} {Phys. Rev. E}\ }\textbf {\bibinfo {volume}
  {104}},\ \bibinfo {pages} {014111} (\bibinfo {year} {2021})}\BibitemShut
  {NoStop}%
\bibitem [{\citenamefont {Torres}\ and\ \citenamefont
  {Salazar-Serrano}(2016)}]{Torres2016}%
  \BibitemOpen
  \bibfield  {author} {\bibinfo {author} {\bibfnamefont {J.~P.}\ \bibnamefont
  {Torres}}\ and\ \bibinfo {author} {\bibfnamefont {L.~J.}\ \bibnamefont
  {Salazar-Serrano}},\ }\href {\doibase 10.1038/srep19702} {\bibfield
  {journal} {\bibinfo  {journal} {Scientific Reports}\ }\textbf {\bibinfo
  {volume} {6}},\ \bibinfo {pages} {19702} (\bibinfo {year}
  {2016})}\BibitemShut {NoStop}%
\bibitem [{\citenamefont {Ren}\ \emph {et~al.}(2020)\citenamefont {Ren},
  \citenamefont {Qin}, \citenamefont {Feng},\ and\ \citenamefont
  {Li}}]{PhysRevA.102.042601}%
  \BibitemOpen
  \bibfield  {author} {\bibinfo {author} {\bibfnamefont {J.}~\bibnamefont
  {Ren}}, \bibinfo {author} {\bibfnamefont {L.}~\bibnamefont {Qin}}, \bibinfo
  {author} {\bibfnamefont {W.}~\bibnamefont {Feng}}, \ and\ \bibinfo {author}
  {\bibfnamefont {X.-Q.}\ \bibnamefont {Li}},\ }\href {\doibase
  10.1103/PhysRevA.102.042601} {\bibfield  {journal} {\bibinfo  {journal}
  {Phys. Rev. A}\ }\textbf {\bibinfo {volume} {102}},\ \bibinfo {pages}
  {042601} (\bibinfo {year} {2020})}\BibitemShut {NoStop}%
\bibitem [{\citenamefont {Dressel}\ \emph {et~al.}(2014)\citenamefont
  {Dressel}, \citenamefont {Malik}, \citenamefont {Miatto}, \citenamefont
  {Jordan},\ and\ \citenamefont {Boyd}}]{RevModPhys.86.307}%
  \BibitemOpen
  \bibfield  {author} {\bibinfo {author} {\bibfnamefont {J.}~\bibnamefont
  {Dressel}}, \bibinfo {author} {\bibfnamefont {M.}~\bibnamefont {Malik}},
  \bibinfo {author} {\bibfnamefont {F.~M.}\ \bibnamefont {Miatto}}, \bibinfo
  {author} {\bibfnamefont {A.~N.}\ \bibnamefont {Jordan}}, \ and\ \bibinfo
  {author} {\bibfnamefont {R.~W.}\ \bibnamefont {Boyd}},\ }\href {\doibase
  10.1103/RevModPhys.86.307} {\bibfield  {journal} {\bibinfo  {journal} {Rev.
  Mod. Phys.}\ }\textbf {\bibinfo {volume} {86}},\ \bibinfo {pages} {307}
  (\bibinfo {year} {2014})}\BibitemShut {NoStop}%
\bibitem [{\citenamefont {Qin}\ \emph {et~al.}(2016)\citenamefont {Qin},
  \citenamefont {Feng},\ and\ \citenamefont {Li}}]{Qin2016}%
  \BibitemOpen
  \bibfield  {author} {\bibinfo {author} {\bibfnamefont {L.}~\bibnamefont
  {Qin}}, \bibinfo {author} {\bibfnamefont {W.}~\bibnamefont {Feng}}, \ and\
  \bibinfo {author} {\bibfnamefont {X.-Q.}\ \bibnamefont {Li}},\ }\href
  {\doibase 10.1038/srep20286} {\bibfield  {journal} {\bibinfo  {journal}
  {Scientific Reports}\ }\textbf {\bibinfo {volume} {6}},\ \bibinfo {pages}
  {20286} (\bibinfo {year} {2016})}\BibitemShut {NoStop}%
\bibitem [{\citenamefont {Vaidman}(2017)}]{doi:10.1098/rsta.2016.0395}%
  \BibitemOpen
  \bibfield  {author} {\bibinfo {author} {\bibfnamefont {L.}~\bibnamefont
  {Vaidman}},\ }\href {\doibase 10.1098/rsta.2016.0395} {\bibfield  {journal}
  {\bibinfo  {journal} {Philosophical Transactions of the Royal Society A:
  Mathematical, Physical and Engineering Sciences}\ }\textbf {\bibinfo {volume}
  {375}},\ \bibinfo {pages} {20160395} (\bibinfo {year} {2017})}\BibitemShut
  {NoStop}%
\bibitem [{\citenamefont {Kunjwal}\ \emph {et~al.}(2019)\citenamefont
  {Kunjwal}, \citenamefont {Lostaglio},\ and\ \citenamefont
  {Pusey}}]{PhysRevA.100.042116}%
  \BibitemOpen
  \bibfield  {author} {\bibinfo {author} {\bibfnamefont {R.}~\bibnamefont
  {Kunjwal}}, \bibinfo {author} {\bibfnamefont {M.}~\bibnamefont {Lostaglio}},
  \ and\ \bibinfo {author} {\bibfnamefont {M.~F.}\ \bibnamefont {Pusey}},\
  }\href {\doibase 10.1103/PhysRevA.100.042116} {\bibfield  {journal} {\bibinfo
   {journal} {Phys. Rev. A}\ }\textbf {\bibinfo {volume} {100}},\ \bibinfo
  {pages} {042116} (\bibinfo {year} {2019})}\BibitemShut {NoStop}%
\bibitem [{\citenamefont {Ogawa}\ \emph {et~al.}(2020)\citenamefont {Ogawa},
  \citenamefont {Kobayashi},\ and\ \citenamefont
  {Tomita}}]{PhysRevA.101.042117}%
  \BibitemOpen
  \bibfield  {author} {\bibinfo {author} {\bibfnamefont {K.}~\bibnamefont
  {Ogawa}}, \bibinfo {author} {\bibfnamefont {H.}~\bibnamefont {Kobayashi}}, \
  and\ \bibinfo {author} {\bibfnamefont {A.}~\bibnamefont {Tomita}},\ }\href
  {\doibase 10.1103/PhysRevA.101.042117} {\bibfield  {journal} {\bibinfo
  {journal} {Phys. Rev. A}\ }\textbf {\bibinfo {volume} {101}},\ \bibinfo
  {pages} {042117} (\bibinfo {year} {2020})}\BibitemShut {NoStop}%
\bibitem [{\citenamefont {Ho}\ and\ \citenamefont
  {Imoto}(2018)}]{PhysRevA.97.012112}%
  \BibitemOpen
  \bibfield  {author} {\bibinfo {author} {\bibfnamefont {L.~B.}\ \bibnamefont
  {Ho}}\ and\ \bibinfo {author} {\bibfnamefont {N.}~\bibnamefont {Imoto}},\
  }\href {\doibase 10.1103/PhysRevA.97.012112} {\bibfield  {journal} {\bibinfo
  {journal} {Phys. Rev. A}\ }\textbf {\bibinfo {volume} {97}},\ \bibinfo
  {pages} {012112} (\bibinfo {year} {2018})}\BibitemShut {NoStop}%
\bibitem [{\citenamefont {Ho}\ and\ \citenamefont {Imoto}(2016)}]{HO20162129}%
  \BibitemOpen
  \bibfield  {author} {\bibinfo {author} {\bibfnamefont {L.~B.}\ \bibnamefont
  {Ho}}\ and\ \bibinfo {author} {\bibfnamefont {N.}~\bibnamefont {Imoto}},\
  }\href {\doibase https://doi.org/10.1016/j.physleta.2016.05.005} {\bibfield
  {journal} {\bibinfo  {journal} {Physics Letters A}\ }\textbf {\bibinfo
  {volume} {380}},\ \bibinfo {pages} {2129 } (\bibinfo {year}
  {2016})}\BibitemShut {NoStop}%
\bibitem [{\citenamefont {Ho}\ and\ \citenamefont
  {Imoto}(2017)}]{PhysRevA.95.032135}%
  \BibitemOpen
  \bibfield  {author} {\bibinfo {author} {\bibfnamefont {L.~B.}\ \bibnamefont
  {Ho}}\ and\ \bibinfo {author} {\bibfnamefont {N.}~\bibnamefont {Imoto}},\
  }\href {\doibase 10.1103/PhysRevA.95.032135} {\bibfield  {journal} {\bibinfo
  {journal} {Phys. Rev. A}\ }\textbf {\bibinfo {volume} {95}},\ \bibinfo
  {pages} {032135} (\bibinfo {year} {2017})}\BibitemShut {NoStop}%
\bibitem [{\citenamefont {Lundeen}\ and\ \citenamefont
  {Steinberg}(2009)}]{PhysRevLett.102.020404}%
  \BibitemOpen
  \bibfield  {author} {\bibinfo {author} {\bibfnamefont {J.~S.}\ \bibnamefont
  {Lundeen}}\ and\ \bibinfo {author} {\bibfnamefont {A.~M.}\ \bibnamefont
  {Steinberg}},\ }\href {\doibase 10.1103/PhysRevLett.102.020404} {\bibfield
  {journal} {\bibinfo  {journal} {Phys. Rev. Lett.}\ }\textbf {\bibinfo
  {volume} {102}},\ \bibinfo {pages} {020404} (\bibinfo {year}
  {2009})}\BibitemShut {NoStop}%
\bibitem [{\citenamefont {Yokota}\ \emph {et~al.}(2009)\citenamefont {Yokota},
  \citenamefont {Yamamoto}, \citenamefont {Koashi},\ and\ \citenamefont
  {Imoto}}]{Yokota_2009}%
  \BibitemOpen
  \bibfield  {author} {\bibinfo {author} {\bibfnamefont {K.}~\bibnamefont
  {Yokota}}, \bibinfo {author} {\bibfnamefont {T.}~\bibnamefont {Yamamoto}},
  \bibinfo {author} {\bibfnamefont {M.}~\bibnamefont {Koashi}}, \ and\ \bibinfo
  {author} {\bibfnamefont {N.}~\bibnamefont {Imoto}},\ }\href {\doibase
  10.1088/1367-2630/11/3/033011} {\bibfield  {journal} {\bibinfo  {journal}
  {New Journal of Physics}\ }\textbf {\bibinfo {volume} {11}},\ \bibinfo
  {pages} {033011} (\bibinfo {year} {2009})}\BibitemShut {NoStop}%
\bibitem [{\citenamefont {Das}\ and\ \citenamefont
  {Sen}(2021)}]{PhysRevA.103.012228}%
  \BibitemOpen
  \bibfield  {author} {\bibinfo {author} {\bibfnamefont {D.}~\bibnamefont
  {Das}}\ and\ \bibinfo {author} {\bibfnamefont {U.}~\bibnamefont {Sen}},\
  }\href {\doibase 10.1103/PhysRevA.103.012228} {\bibfield  {journal} {\bibinfo
   {journal} {Phys. Rev. A}\ }\textbf {\bibinfo {volume} {103}},\ \bibinfo
  {pages} {012228} (\bibinfo {year} {2021})}\BibitemShut {NoStop}%
\bibitem [{\citenamefont {Calder{\'o}n-Losada}\ \emph
  {et~al.}(2020)\citenamefont {Calder{\'o}n-Losada}, \citenamefont
  {Moctezuma~Quistian}, \citenamefont {Cruz-Ramirez}, \citenamefont
  {Murgueitio~Ramirez}, \citenamefont {U'Ren}, \citenamefont {Botero},\ and\
  \citenamefont {Valencia}}]{Calderon-Losada2020}%
  \BibitemOpen
  \bibfield  {author} {\bibinfo {author} {\bibfnamefont {O.}~\bibnamefont
  {Calder{\'o}n-Losada}}, \bibinfo {author} {\bibfnamefont {T.~T.}\
  \bibnamefont {Moctezuma~Quistian}}, \bibinfo {author} {\bibfnamefont
  {H.}~\bibnamefont {Cruz-Ramirez}}, \bibinfo {author} {\bibfnamefont
  {S.}~\bibnamefont {Murgueitio~Ramirez}}, \bibinfo {author} {\bibfnamefont
  {A.~B.}\ \bibnamefont {U'Ren}}, \bibinfo {author} {\bibfnamefont
  {A.}~\bibnamefont {Botero}}, \ and\ \bibinfo {author} {\bibfnamefont
  {A.}~\bibnamefont {Valencia}},\ }\href {\doibase 10.1038/s42005-020-0378-3}
  {\bibfield  {journal} {\bibinfo  {journal} {Communications Physics}\ }\textbf
  {\bibinfo {volume} {3}},\ \bibinfo {pages} {117} (\bibinfo {year}
  {2020})}\BibitemShut {NoStop}%
\bibitem [{\citenamefont {Xu}\ \emph {et~al.}(2020)\citenamefont {Xu},
  \citenamefont {Pan}, \citenamefont {Kedem}, \citenamefont {Wang},
  \citenamefont {Sun}, \citenamefont {Xu}, \citenamefont {Han}, \citenamefont
  {Chen}, \citenamefont {Li},\ and\ \citenamefont {Guo}}]{Xu:20}%
  \BibitemOpen
  \bibfield  {author} {\bibinfo {author} {\bibfnamefont {X.-Y.}\ \bibnamefont
  {Xu}}, \bibinfo {author} {\bibfnamefont {W.-W.}\ \bibnamefont {Pan}},
  \bibinfo {author} {\bibfnamefont {Y.}~\bibnamefont {Kedem}}, \bibinfo
  {author} {\bibfnamefont {Q.-Q.}\ \bibnamefont {Wang}}, \bibinfo {author}
  {\bibfnamefont {K.}~\bibnamefont {Sun}}, \bibinfo {author} {\bibfnamefont
  {J.-S.}\ \bibnamefont {Xu}}, \bibinfo {author} {\bibfnamefont {Y.-J.}\
  \bibnamefont {Han}}, \bibinfo {author} {\bibfnamefont {G.}~\bibnamefont
  {Chen}}, \bibinfo {author} {\bibfnamefont {C.-F.}\ \bibnamefont {Li}}, \ and\
  \bibinfo {author} {\bibfnamefont {G.-C.}\ \bibnamefont {Guo}},\ }\href
  {\doibase 10.1364/OL.375448} {\bibfield  {journal} {\bibinfo  {journal} {Opt.
  Lett.}\ }\textbf {\bibinfo {volume} {45}},\ \bibinfo {pages} {1715} (\bibinfo
  {year} {2020})}\BibitemShut {NoStop}%
\bibitem [{\citenamefont {Aharonov}\ \emph {et~al.}(2013)\citenamefont
  {Aharonov}, \citenamefont {Popescu}, \citenamefont {Rohrlich},\ and\
  \citenamefont {Skrzypczyk}}]{Aharonov_2013}%
  \BibitemOpen
  \bibfield  {author} {\bibinfo {author} {\bibfnamefont {Y.}~\bibnamefont
  {Aharonov}}, \bibinfo {author} {\bibfnamefont {S.}~\bibnamefont {Popescu}},
  \bibinfo {author} {\bibfnamefont {D.}~\bibnamefont {Rohrlich}}, \ and\
  \bibinfo {author} {\bibfnamefont {P.}~\bibnamefont {Skrzypczyk}},\ }\href
  {\doibase 10.1088/1367-2630/15/11/113015} {\bibfield  {journal} {\bibinfo
  {journal} {New Journal of Physics}\ }\textbf {\bibinfo {volume} {15}},\
  \bibinfo {pages} {113015} (\bibinfo {year} {2013})}\BibitemShut {NoStop}%
\bibitem [{\citenamefont {Denkmayr}\ \emph {et~al.}(2014)\citenamefont
  {Denkmayr}, \citenamefont {Geppert}, \citenamefont {Sponar}, \citenamefont
  {Lemmel}, \citenamefont {Matzkin}, \citenamefont {Tollaksen},\ and\
  \citenamefont {Hasegawa}}]{Denkmayr2014}%
  \BibitemOpen
  \bibfield  {author} {\bibinfo {author} {\bibfnamefont {T.}~\bibnamefont
  {Denkmayr}}, \bibinfo {author} {\bibfnamefont {H.}~\bibnamefont {Geppert}},
  \bibinfo {author} {\bibfnamefont {S.}~\bibnamefont {Sponar}}, \bibinfo
  {author} {\bibfnamefont {H.}~\bibnamefont {Lemmel}}, \bibinfo {author}
  {\bibfnamefont {A.}~\bibnamefont {Matzkin}}, \bibinfo {author} {\bibfnamefont
  {J.}~\bibnamefont {Tollaksen}}, \ and\ \bibinfo {author} {\bibfnamefont
  {Y.}~\bibnamefont {Hasegawa}},\ }\href {\doibase 10.1038/ncomms5492}
  {\bibfield  {journal} {\bibinfo  {journal} {Nature Communications}\ }\textbf
  {\bibinfo {volume} {5}},\ \bibinfo {pages} {4492} (\bibinfo {year}
  {2014})}\BibitemShut {NoStop}%
\bibitem [{\citenamefont {Kim}\ \emph {et~al.}(2021)\citenamefont {Kim},
  \citenamefont {Im}, \citenamefont {Kim}, \citenamefont {Han}, \citenamefont
  {Moon}, \citenamefont {Kim},\ and\ \citenamefont {Cho}}]{Kim2021}%
  \BibitemOpen
  \bibfield  {author} {\bibinfo {author} {\bibfnamefont {Y.}~\bibnamefont
  {Kim}}, \bibinfo {author} {\bibfnamefont {D.-G.}\ \bibnamefont {Im}},
  \bibinfo {author} {\bibfnamefont {Y.-S.}\ \bibnamefont {Kim}}, \bibinfo
  {author} {\bibfnamefont {S.-W.}\ \bibnamefont {Han}}, \bibinfo {author}
  {\bibfnamefont {S.}~\bibnamefont {Moon}}, \bibinfo {author} {\bibfnamefont
  {Y.-H.}\ \bibnamefont {Kim}}, \ and\ \bibinfo {author} {\bibfnamefont
  {Y.-W.}\ \bibnamefont {Cho}},\ }\href {\doibase 10.1038/s41534-020-00350-6}
  {\bibfield  {journal} {\bibinfo  {journal} {npj Quantum Information}\
  }\textbf {\bibinfo {volume} {7}},\ \bibinfo {pages} {13} (\bibinfo {year}
  {2021})}\BibitemShut {NoStop}%
\bibitem [{\citenamefont {Aharonov}\ \emph {et~al.}(2016)\citenamefont
  {Aharonov}, \citenamefont {Colombo}, \citenamefont {Popescu}, \citenamefont
  {Sabadini}, \citenamefont {Struppa},\ and\ \citenamefont
  {Tollaksen}}]{Aharonov532}%
  \BibitemOpen
  \bibfield  {author} {\bibinfo {author} {\bibfnamefont {Y.}~\bibnamefont
  {Aharonov}}, \bibinfo {author} {\bibfnamefont {F.}~\bibnamefont {Colombo}},
  \bibinfo {author} {\bibfnamefont {S.}~\bibnamefont {Popescu}}, \bibinfo
  {author} {\bibfnamefont {I.}~\bibnamefont {Sabadini}}, \bibinfo {author}
  {\bibfnamefont {D.~C.}\ \bibnamefont {Struppa}}, \ and\ \bibinfo {author}
  {\bibfnamefont {J.}~\bibnamefont {Tollaksen}},\ }\href {\doibase
  10.1073/pnas.1522411112} {\bibfield  {journal} {\bibinfo  {journal}
  {Proceedings of the National Academy of Sciences}\ }\textbf {\bibinfo
  {volume} {113}},\ \bibinfo {pages} {532} (\bibinfo {year}
  {2016})}\BibitemShut {NoStop}%
\bibitem [{\citenamefont {Waegell}\ \emph {et~al.}(2017)\citenamefont
  {Waegell}, \citenamefont {Denkmayr}, \citenamefont {Geppert}, \citenamefont
  {Ebner}, \citenamefont {Jenke}, \citenamefont {Hasegawa}, \citenamefont
  {Sponar}, \citenamefont {Dressel},\ and\ \citenamefont
  {Tollaksen}}]{PhysRevA.96.052131}%
  \BibitemOpen
  \bibfield  {author} {\bibinfo {author} {\bibfnamefont {M.}~\bibnamefont
  {Waegell}}, \bibinfo {author} {\bibfnamefont {T.}~\bibnamefont {Denkmayr}},
  \bibinfo {author} {\bibfnamefont {H.}~\bibnamefont {Geppert}}, \bibinfo
  {author} {\bibfnamefont {D.}~\bibnamefont {Ebner}}, \bibinfo {author}
  {\bibfnamefont {T.}~\bibnamefont {Jenke}}, \bibinfo {author} {\bibfnamefont
  {Y.}~\bibnamefont {Hasegawa}}, \bibinfo {author} {\bibfnamefont
  {S.}~\bibnamefont {Sponar}}, \bibinfo {author} {\bibfnamefont
  {J.}~\bibnamefont {Dressel}}, \ and\ \bibinfo {author} {\bibfnamefont
  {J.}~\bibnamefont {Tollaksen}},\ }\href {\doibase 10.1103/PhysRevA.96.052131}
  {\bibfield  {journal} {\bibinfo  {journal} {Phys. Rev. A}\ }\textbf {\bibinfo
  {volume} {96}},\ \bibinfo {pages} {052131} (\bibinfo {year}
  {2017})}\BibitemShut {NoStop}%
\bibitem [{\citenamefont {Budiyono}\ and\ \citenamefont
  {Dipojono}(2021)}]{PhysRevA.103.022215}%
  \BibitemOpen
  \bibfield  {author} {\bibinfo {author} {\bibfnamefont {A.}~\bibnamefont
  {Budiyono}}\ and\ \bibinfo {author} {\bibfnamefont {H.~K.}\ \bibnamefont
  {Dipojono}},\ }\href {\doibase 10.1103/PhysRevA.103.022215} {\bibfield
  {journal} {\bibinfo  {journal} {Phys. Rev. A}\ }\textbf {\bibinfo {volume}
  {103}},\ \bibinfo {pages} {022215} (\bibinfo {year} {2021})}\BibitemShut
  {NoStop}%
\bibitem [{\citenamefont {Hosten}\ and\ \citenamefont
  {Kwiat}(2008)}]{Hosten787}%
  \BibitemOpen
  \bibfield  {author} {\bibinfo {author} {\bibfnamefont {O.}~\bibnamefont
  {Hosten}}\ and\ \bibinfo {author} {\bibfnamefont {P.}~\bibnamefont {Kwiat}},\
  }\href {\doibase 10.1126/science.1152697} {\bibfield  {journal} {\bibinfo
  {journal} {Science}\ }\textbf {\bibinfo {volume} {319}},\ \bibinfo {pages}
  {787} (\bibinfo {year} {2008})}\BibitemShut {NoStop}%
\bibitem [{\citenamefont {Dixon}\ \emph {et~al.}(2009)\citenamefont {Dixon},
  \citenamefont {Starling}, \citenamefont {Jordan},\ and\ \citenamefont
  {Howell}}]{PhysRevLett.102.173601}%
  \BibitemOpen
  \bibfield  {author} {\bibinfo {author} {\bibfnamefont {P.~B.}\ \bibnamefont
  {Dixon}}, \bibinfo {author} {\bibfnamefont {D.~J.}\ \bibnamefont {Starling}},
  \bibinfo {author} {\bibfnamefont {A.~N.}\ \bibnamefont {Jordan}}, \ and\
  \bibinfo {author} {\bibfnamefont {J.~C.}\ \bibnamefont {Howell}},\ }\href
  {\doibase 10.1103/PhysRevLett.102.173601} {\bibfield  {journal} {\bibinfo
  {journal} {Phys. Rev. Lett.}\ }\textbf {\bibinfo {volume} {102}},\ \bibinfo
  {pages} {173601} (\bibinfo {year} {2009})}\BibitemShut {NoStop}%
\bibitem [{\citenamefont {Modak}\ \emph {et~al.}(2021)\citenamefont {Modak},
  \citenamefont {B~S}, \citenamefont {Singh},\ and\ \citenamefont
  {Ghosh}}]{PhysRevA.103.053518}%
  \BibitemOpen
  \bibfield  {author} {\bibinfo {author} {\bibfnamefont {N.}~\bibnamefont
  {Modak}}, \bibinfo {author} {\bibfnamefont {A.}~\bibnamefont {B~S}}, \bibinfo
  {author} {\bibfnamefont {A.~K.}\ \bibnamefont {Singh}}, \ and\ \bibinfo
  {author} {\bibfnamefont {N.}~\bibnamefont {Ghosh}},\ }\href {\doibase
  10.1103/PhysRevA.103.053518} {\bibfield  {journal} {\bibinfo  {journal}
  {Phys. Rev. A}\ }\textbf {\bibinfo {volume} {103}},\ \bibinfo {pages}
  {053518} (\bibinfo {year} {2021})}\BibitemShut {NoStop}%
\bibitem [{\citenamefont {Krafczyk}\ \emph {et~al.}(2021)\citenamefont
  {Krafczyk}, \citenamefont {Jordan}, \citenamefont {Goggin},\ and\
  \citenamefont {Kwiat}}]{PhysRevLett.126.220801}%
  \BibitemOpen
  \bibfield  {author} {\bibinfo {author} {\bibfnamefont {C.}~\bibnamefont
  {Krafczyk}}, \bibinfo {author} {\bibfnamefont {A.~N.}\ \bibnamefont
  {Jordan}}, \bibinfo {author} {\bibfnamefont {M.~E.}\ \bibnamefont {Goggin}},
  \ and\ \bibinfo {author} {\bibfnamefont {P.~G.}\ \bibnamefont {Kwiat}},\
  }\href {\doibase 10.1103/PhysRevLett.126.220801} {\bibfield  {journal}
  {\bibinfo  {journal} {Phys. Rev. Lett.}\ }\textbf {\bibinfo {volume} {126}},\
  \bibinfo {pages} {220801} (\bibinfo {year} {2021})}\BibitemShut {NoStop}%
\bibitem [{\citenamefont {Zhou}\ \emph {et~al.}(2021)\citenamefont {Zhou},
  \citenamefont {Cheng}, \citenamefont {Liu}, \citenamefont {Zhang},
  \citenamefont {Yang},\ and\ \citenamefont {Luo}}]{ZHOU2021126655}%
  \BibitemOpen
  \bibfield  {author} {\bibinfo {author} {\bibfnamefont {X.}~\bibnamefont
  {Zhou}}, \bibinfo {author} {\bibfnamefont {W.}~\bibnamefont {Cheng}},
  \bibinfo {author} {\bibfnamefont {S.}~\bibnamefont {Liu}}, \bibinfo {author}
  {\bibfnamefont {J.}~\bibnamefont {Zhang}}, \bibinfo {author} {\bibfnamefont
  {C.}~\bibnamefont {Yang}}, \ and\ \bibinfo {author} {\bibfnamefont
  {Z.}~\bibnamefont {Luo}},\ }\href {\doibase
  https://doi.org/10.1016/j.optcom.2020.126655} {\bibfield  {journal} {\bibinfo
   {journal} {Optics Communications}\ }\textbf {\bibinfo {volume} {483}},\
  \bibinfo {pages} {126655} (\bibinfo {year} {2021})}\BibitemShut {NoStop}%
\bibitem [{\citenamefont {Fang}\ \emph {et~al.}(2021)\citenamefont {Fang},
  \citenamefont {Xia}, \citenamefont {Huang}, \citenamefont {Xiao},
  \citenamefont {Yu}, \citenamefont {Li},\ and\ \citenamefont
  {Zeng}}]{Fang_2021}%
  \BibitemOpen
  \bibfield  {author} {\bibinfo {author} {\bibfnamefont {C.}~\bibnamefont
  {Fang}}, \bibinfo {author} {\bibfnamefont {B.}~\bibnamefont {Xia}}, \bibinfo
  {author} {\bibfnamefont {J.-Z.}\ \bibnamefont {Huang}}, \bibinfo {author}
  {\bibfnamefont {T.}~\bibnamefont {Xiao}}, \bibinfo {author} {\bibfnamefont
  {Y.}~\bibnamefont {Yu}}, \bibinfo {author} {\bibfnamefont {H.}~\bibnamefont
  {Li}}, \ and\ \bibinfo {author} {\bibfnamefont {G.}~\bibnamefont {Zeng}},\
  }\href {\doibase 10.1088/1361-6455/abe5c7} {\bibfield  {journal} {\bibinfo
  {journal} {Journal of Physics B: Atomic, Molecular and Optical Physics}\
  }\textbf {\bibinfo {volume} {54}},\ \bibinfo {pages} {075501} (\bibinfo
  {year} {2021})}\BibitemShut {NoStop}%
\bibitem [{\citenamefont {Liu}\ \emph {et~al.}(2021)\citenamefont {Liu},
  \citenamefont {Li}, \citenamefont {Wang}, \citenamefont {Xia}, \citenamefont
  {Huang},\ and\ \citenamefont {Zeng}}]{Liu_2021}%
  \BibitemOpen
  \bibfield  {author} {\bibinfo {author} {\bibfnamefont {M.}~\bibnamefont
  {Liu}}, \bibinfo {author} {\bibfnamefont {H.}~\bibnamefont {Li}}, \bibinfo
  {author} {\bibfnamefont {G.}~\bibnamefont {Wang}}, \bibinfo {author}
  {\bibfnamefont {B.}~\bibnamefont {Xia}}, \bibinfo {author} {\bibfnamefont
  {J.}~\bibnamefont {Huang}}, \ and\ \bibinfo {author} {\bibfnamefont
  {G.}~\bibnamefont {Zeng}},\ }\href {\doibase 10.1088/1361-6455/abc59f}
  {\bibfield  {journal} {\bibinfo  {journal} {Journal of Physics B: Atomic,
  Molecular and Optical Physics}\ }\textbf {\bibinfo {volume} {54}},\ \bibinfo
  {pages} {085501} (\bibinfo {year} {2021})}\BibitemShut {NoStop}%
\bibitem [{\citenamefont {Huang}\ \emph {et~al.}(2021)\citenamefont {Huang},
  \citenamefont {Duan},\ and\ \citenamefont {Hu}}]{Huang2021}%
  \BibitemOpen
  \bibfield  {author} {\bibinfo {author} {\bibfnamefont {J.-H.}\ \bibnamefont
  {Huang}}, \bibinfo {author} {\bibfnamefont {X.-Y.}\ \bibnamefont {Duan}}, \
  and\ \bibinfo {author} {\bibfnamefont {X.-Y.}\ \bibnamefont {Hu}},\ }\href
  {\doibase 10.1140/epjd/s10053-021-00117-4} {\bibfield  {journal} {\bibinfo
  {journal} {The European Physical Journal D}\ }\textbf {\bibinfo {volume}
  {75}},\ \bibinfo {pages} {114} (\bibinfo {year} {2021})}\BibitemShut
  {NoStop}%
\bibitem [{\citenamefont {Arvidsson-Shukur}\ \emph {et~al.}(2020)\citenamefont
  {Arvidsson-Shukur}, \citenamefont {Yunger~Halpern}, \citenamefont {Lepage},
  \citenamefont {Lasek}, \citenamefont {Barnes},\ and\ \citenamefont
  {Lloyd}}]{Arvidsson-Shukur2020}%
  \BibitemOpen
  \bibfield  {author} {\bibinfo {author} {\bibfnamefont {D.~R.~M.}\
  \bibnamefont {Arvidsson-Shukur}}, \bibinfo {author} {\bibfnamefont
  {N.}~\bibnamefont {Yunger~Halpern}}, \bibinfo {author} {\bibfnamefont
  {H.~V.}\ \bibnamefont {Lepage}}, \bibinfo {author} {\bibfnamefont {A.~A.}\
  \bibnamefont {Lasek}}, \bibinfo {author} {\bibfnamefont {C.~H.~W.}\
  \bibnamefont {Barnes}}, \ and\ \bibinfo {author} {\bibfnamefont
  {S.}~\bibnamefont {Lloyd}},\ }\href {\doibase 10.1038/s41467-020-17559-w}
  {\bibfield  {journal} {\bibinfo  {journal} {Nature Communications}\ }\textbf
  {\bibinfo {volume} {11}},\ \bibinfo {pages} {3775} (\bibinfo {year}
  {2020})}\BibitemShut {NoStop}%
\bibitem [{\citenamefont {Pati}\ \emph {et~al.}(2020)\citenamefont {Pati},
  \citenamefont {Mukhopadhyay}, \citenamefont {Chakraborty},\ and\
  \citenamefont {Ghosh}}]{PhysRevA.102.012204}%
  \BibitemOpen
  \bibfield  {author} {\bibinfo {author} {\bibfnamefont {A.~K.}\ \bibnamefont
  {Pati}}, \bibinfo {author} {\bibfnamefont {C.}~\bibnamefont {Mukhopadhyay}},
  \bibinfo {author} {\bibfnamefont {S.}~\bibnamefont {Chakraborty}}, \ and\
  \bibinfo {author} {\bibfnamefont {S.}~\bibnamefont {Ghosh}},\ }\href
  {\doibase 10.1103/PhysRevA.102.012204} {\bibfield  {journal} {\bibinfo
  {journal} {Phys. Rev. A}\ }\textbf {\bibinfo {volume} {102}},\ \bibinfo
  {pages} {012204} (\bibinfo {year} {2020})}\BibitemShut {NoStop}%
\bibitem [{\citenamefont {Yin}\ \emph {et~al.}(2021)\citenamefont {Yin},
  \citenamefont {Zhang}, \citenamefont {Xu}, \citenamefont {Liu}, \citenamefont
  {Zhuang}, \citenamefont {Chen}, \citenamefont {Gong}, \citenamefont {Ma},
  \citenamefont {Peng}, \citenamefont {Li}, \citenamefont {Xu}, \citenamefont
  {Zhou}, \citenamefont {Zhang}, \citenamefont {Chen}, \citenamefont {Li},\
  and\ \citenamefont {Guo}}]{Yin2021}%
  \BibitemOpen
  \bibfield  {author} {\bibinfo {author} {\bibfnamefont {P.}~\bibnamefont
  {Yin}}, \bibinfo {author} {\bibfnamefont {W.-H.}\ \bibnamefont {Zhang}},
  \bibinfo {author} {\bibfnamefont {L.}~\bibnamefont {Xu}}, \bibinfo {author}
  {\bibfnamefont {Z.-G.}\ \bibnamefont {Liu}}, \bibinfo {author} {\bibfnamefont
  {W.-F.}\ \bibnamefont {Zhuang}}, \bibinfo {author} {\bibfnamefont
  {L.}~\bibnamefont {Chen}}, \bibinfo {author} {\bibfnamefont {M.}~\bibnamefont
  {Gong}}, \bibinfo {author} {\bibfnamefont {Y.}~\bibnamefont {Ma}}, \bibinfo
  {author} {\bibfnamefont {X.-X.}\ \bibnamefont {Peng}}, \bibinfo {author}
  {\bibfnamefont {G.-C.}\ \bibnamefont {Li}}, \bibinfo {author} {\bibfnamefont
  {J.-S.}\ \bibnamefont {Xu}}, \bibinfo {author} {\bibfnamefont {Z.-Q.}\
  \bibnamefont {Zhou}}, \bibinfo {author} {\bibfnamefont {L.}~\bibnamefont
  {Zhang}}, \bibinfo {author} {\bibfnamefont {G.}~\bibnamefont {Chen}},
  \bibinfo {author} {\bibfnamefont {C.-F.}\ \bibnamefont {Li}}, \ and\ \bibinfo
  {author} {\bibfnamefont {G.-C.}\ \bibnamefont {Guo}},\ }\href {\doibase
  10.1038/s41377-021-00543-4} {\bibfield  {journal} {\bibinfo  {journal}
  {Light: Science {\&} Applications}\ }\textbf {\bibinfo {volume} {10}},\
  \bibinfo {pages} {103} (\bibinfo {year} {2021})}\BibitemShut {NoStop}%
\bibitem [{\citenamefont {Ho}\ and\ \citenamefont
  {Kondo}(2021)}]{doi:10.1063/5.0024555}%
  \BibitemOpen
  \bibfield  {author} {\bibinfo {author} {\bibfnamefont {L.~B.}\ \bibnamefont
  {Ho}}\ and\ \bibinfo {author} {\bibfnamefont {Y.}~\bibnamefont {Kondo}},\
  }\href {\doibase 10.1063/5.0024555} {\bibfield  {journal} {\bibinfo
  {journal} {Journal of Mathematical Physics}\ }\textbf {\bibinfo {volume}
  {62}},\ \bibinfo {pages} {012102} (\bibinfo {year} {2021})}\BibitemShut
  {NoStop}%
\bibitem [{\citenamefont {Ho}\ and\ \citenamefont {Kondo}(2019)}]{HO2019153}%
  \BibitemOpen
  \bibfield  {author} {\bibinfo {author} {\bibfnamefont {L.~B.}\ \bibnamefont
  {Ho}}\ and\ \bibinfo {author} {\bibfnamefont {Y.}~\bibnamefont {Kondo}},\
  }\href {\doibase https://doi.org/10.1016/j.physleta.2018.10.041} {\bibfield
  {journal} {\bibinfo  {journal} {Physics Letters A}\ }\textbf {\bibinfo
  {volume} {383}},\ \bibinfo {pages} {153} (\bibinfo {year}
  {2019})}\BibitemShut {NoStop}%
\bibitem [{\citenamefont {Harris}\ \emph {et~al.}(2017)\citenamefont {Harris},
  \citenamefont {Boyd},\ and\ \citenamefont
  {Lundeen}}]{PhysRevLett.118.070802}%
  \BibitemOpen
  \bibfield  {author} {\bibinfo {author} {\bibfnamefont {J.}~\bibnamefont
  {Harris}}, \bibinfo {author} {\bibfnamefont {R.~W.}\ \bibnamefont {Boyd}}, \
  and\ \bibinfo {author} {\bibfnamefont {J.~S.}\ \bibnamefont {Lundeen}},\
  }\href {\doibase 10.1103/PhysRevLett.118.070802} {\bibfield  {journal}
  {\bibinfo  {journal} {Phys. Rev. Lett.}\ }\textbf {\bibinfo {volume} {118}},\
  \bibinfo {pages} {070802} (\bibinfo {year} {2017})}\BibitemShut {NoStop}%
\bibitem [{\citenamefont {Pang}\ and\ \citenamefont
  {Brun}(2015)}]{PhysRevLett.115.120401}%
  \BibitemOpen
  \bibfield  {author} {\bibinfo {author} {\bibfnamefont {S.}~\bibnamefont
  {Pang}}\ and\ \bibinfo {author} {\bibfnamefont {T.~A.}\ \bibnamefont
  {Brun}},\ }\href {\doibase 10.1103/PhysRevLett.115.120401} {\bibfield
  {journal} {\bibinfo  {journal} {Phys. Rev. Lett.}\ }\textbf {\bibinfo
  {volume} {115}},\ \bibinfo {pages} {120401} (\bibinfo {year}
  {2015})}\BibitemShut {NoStop}%
\bibitem [{\citenamefont {Lundeen}\ \emph {et~al.}(2011)\citenamefont
  {Lundeen}, \citenamefont {Sutherland}, \citenamefont {Patel}, \citenamefont
  {Stewart},\ and\ \citenamefont {Bamber}}]{Lundeen2011}%
  \BibitemOpen
  \bibfield  {author} {\bibinfo {author} {\bibfnamefont {J.~S.}\ \bibnamefont
  {Lundeen}}, \bibinfo {author} {\bibfnamefont {B.}~\bibnamefont {Sutherland}},
  \bibinfo {author} {\bibfnamefont {A.}~\bibnamefont {Patel}}, \bibinfo
  {author} {\bibfnamefont {C.}~\bibnamefont {Stewart}}, \ and\ \bibinfo
  {author} {\bibfnamefont {C.}~\bibnamefont {Bamber}},\ }\href {\doibase
  10.1038/nature10120} {\bibfield  {journal} {\bibinfo  {journal} {Nature}\
  }\textbf {\bibinfo {volume} {474}},\ \bibinfo {pages} {188} (\bibinfo {year}
  {2011})}\BibitemShut {NoStop}%
\bibitem [{\citenamefont {Lundeen}\ and\ \citenamefont
  {Bamber}(2012)}]{PhysRevLett.108.070402}%
  \BibitemOpen
  \bibfield  {author} {\bibinfo {author} {\bibfnamefont {J.~S.}\ \bibnamefont
  {Lundeen}}\ and\ \bibinfo {author} {\bibfnamefont {C.}~\bibnamefont
  {Bamber}},\ }\href {\doibase 10.1103/PhysRevLett.108.070402} {\bibfield
  {journal} {\bibinfo  {journal} {Phys. Rev. Lett.}\ }\textbf {\bibinfo
  {volume} {108}},\ \bibinfo {pages} {070402} (\bibinfo {year}
  {2012})}\BibitemShut {NoStop}%
\bibitem [{\citenamefont {Calderaro}\ \emph {et~al.}(2018)\citenamefont
  {Calderaro}, \citenamefont {Foletto}, \citenamefont {Dequal}, \citenamefont
  {Villoresi},\ and\ \citenamefont {Vallone}}]{PhysRevLett.121.230501}%
  \BibitemOpen
  \bibfield  {author} {\bibinfo {author} {\bibfnamefont {L.}~\bibnamefont
  {Calderaro}}, \bibinfo {author} {\bibfnamefont {G.}~\bibnamefont {Foletto}},
  \bibinfo {author} {\bibfnamefont {D.}~\bibnamefont {Dequal}}, \bibinfo
  {author} {\bibfnamefont {P.}~\bibnamefont {Villoresi}}, \ and\ \bibinfo
  {author} {\bibfnamefont {G.}~\bibnamefont {Vallone}},\ }\href {\doibase
  10.1103/PhysRevLett.121.230501} {\bibfield  {journal} {\bibinfo  {journal}
  {Phys. Rev. Lett.}\ }\textbf {\bibinfo {volume} {121}},\ \bibinfo {pages}
  {230501} (\bibinfo {year} {2018})}\BibitemShut {NoStop}%
\bibitem [{\citenamefont {Vallone}\ and\ \citenamefont
  {Dequal}(2016)}]{PhysRevLett.116.040502}%
  \BibitemOpen
  \bibfield  {author} {\bibinfo {author} {\bibfnamefont {G.}~\bibnamefont
  {Vallone}}\ and\ \bibinfo {author} {\bibfnamefont {D.}~\bibnamefont
  {Dequal}},\ }\href {\doibase 10.1103/PhysRevLett.116.040502} {\bibfield
  {journal} {\bibinfo  {journal} {Phys. Rev. Lett.}\ }\textbf {\bibinfo
  {volume} {116}},\ \bibinfo {pages} {040502} (\bibinfo {year}
  {2016})}\BibitemShut {NoStop}%
\bibitem [{\citenamefont {Maccone}\ and\ \citenamefont
  {Rusconi}(2014)}]{PhysRevA.89.022122}%
  \BibitemOpen
  \bibfield  {author} {\bibinfo {author} {\bibfnamefont {L.}~\bibnamefont
  {Maccone}}\ and\ \bibinfo {author} {\bibfnamefont {C.~C.}\ \bibnamefont
  {Rusconi}},\ }\href {\doibase 10.1103/PhysRevA.89.022122} {\bibfield
  {journal} {\bibinfo  {journal} {Phys. Rev. A}\ }\textbf {\bibinfo {volume}
  {89}},\ \bibinfo {pages} {022122} (\bibinfo {year} {2014})}\BibitemShut
  {NoStop}%
\bibitem [{\citenamefont {Chen}\ \emph {et~al.}(2018)\citenamefont {Chen},
  \citenamefont {Dai}, \citenamefont {Yang},\ and\ \citenamefont
  {Zhang}}]{PhysRevA.97.032120}%
  \BibitemOpen
  \bibfield  {author} {\bibinfo {author} {\bibfnamefont {X.}~\bibnamefont
  {Chen}}, \bibinfo {author} {\bibfnamefont {H.-Y.}\ \bibnamefont {Dai}},
  \bibinfo {author} {\bibfnamefont {L.}~\bibnamefont {Yang}}, \ and\ \bibinfo
  {author} {\bibfnamefont {M.}~\bibnamefont {Zhang}},\ }\href {\doibase
  10.1103/PhysRevA.97.032120} {\bibfield  {journal} {\bibinfo  {journal} {Phys.
  Rev. A}\ }\textbf {\bibinfo {volume} {97}},\ \bibinfo {pages} {032120}
  (\bibinfo {year} {2018})}\BibitemShut {NoStop}%
\bibitem [{\citenamefont {Pan}\ \emph {et~al.}(2019)\citenamefont {Pan},
  \citenamefont {Xu}, \citenamefont {Kedem}, \citenamefont {Wang},
  \citenamefont {Chen}, \citenamefont {Jan}, \citenamefont {Sun}, \citenamefont
  {Xu}, \citenamefont {Han}, \citenamefont {Li},\ and\ \citenamefont
  {Guo}}]{PhysRevLett.123.150402}%
  \BibitemOpen
  \bibfield  {author} {\bibinfo {author} {\bibfnamefont {W.-W.}\ \bibnamefont
  {Pan}}, \bibinfo {author} {\bibfnamefont {X.-Y.}\ \bibnamefont {Xu}},
  \bibinfo {author} {\bibfnamefont {Y.}~\bibnamefont {Kedem}}, \bibinfo
  {author} {\bibfnamefont {Q.-Q.}\ \bibnamefont {Wang}}, \bibinfo {author}
  {\bibfnamefont {Z.}~\bibnamefont {Chen}}, \bibinfo {author} {\bibfnamefont
  {M.}~\bibnamefont {Jan}}, \bibinfo {author} {\bibfnamefont {K.}~\bibnamefont
  {Sun}}, \bibinfo {author} {\bibfnamefont {J.-S.}\ \bibnamefont {Xu}},
  \bibinfo {author} {\bibfnamefont {Y.-J.}\ \bibnamefont {Han}}, \bibinfo
  {author} {\bibfnamefont {C.-F.}\ \bibnamefont {Li}}, \ and\ \bibinfo {author}
  {\bibfnamefont {G.-C.}\ \bibnamefont {Guo}},\ }\href {\doibase
  10.1103/PhysRevLett.123.150402} {\bibfield  {journal} {\bibinfo  {journal}
  {Phys. Rev. Lett.}\ }\textbf {\bibinfo {volume} {123}},\ \bibinfo {pages}
  {150402} (\bibinfo {year} {2019})}\BibitemShut {NoStop}%
\bibitem [{\citenamefont {Turek}(2020)}]{Turek_2020}%
  \BibitemOpen
  \bibfield  {author} {\bibinfo {author} {\bibfnamefont {Y.}~\bibnamefont
  {Turek}},\ }\href {\doibase 10.1088/2399-6528/ab9938} {\bibfield  {journal}
  {\bibinfo  {journal} {Journal of Physics Communications}\ }\textbf {\bibinfo
  {volume} {4}},\ \bibinfo {pages} {075007} (\bibinfo {year}
  {2020})}\BibitemShut {NoStop}%
\bibitem [{\citenamefont {Ho}(2020)}]{Ho_2020}%
  \BibitemOpen
  \bibfield  {author} {\bibinfo {author} {\bibfnamefont {L.~B.}\ \bibnamefont
  {Ho}},\ }\href {\doibase 10.1088/1361-6455/ab7881} {\bibfield  {journal}
  {\bibinfo  {journal} {Journal of Physics B: Atomic, Molecular and Optical
  Physics}\ }\textbf {\bibinfo {volume} {53}},\ \bibinfo {pages} {115501}
  (\bibinfo {year} {2020})}\BibitemShut {NoStop}%
\bibitem [{\citenamefont {Tuan}\ \emph {et~al.}(2021)\citenamefont {Tuan},
  \citenamefont {Nguyen},\ and\ \citenamefont {Ho}}]{Tuan2021}%
  \BibitemOpen
  \bibfield  {author} {\bibinfo {author} {\bibfnamefont {K.~Q.}\ \bibnamefont
  {Tuan}}, \bibinfo {author} {\bibfnamefont {H.~Q.}\ \bibnamefont {Nguyen}}, \
  and\ \bibinfo {author} {\bibfnamefont {L.~B.}\ \bibnamefont {Ho}},\ }\href
  {\doibase 10.1007/s11128-021-03144-7} {\bibfield  {journal} {\bibinfo
  {journal} {Quantum Information Processing}\ }\textbf {\bibinfo {volume}
  {20}},\ \bibinfo {pages} {197} (\bibinfo {year} {2021})}\BibitemShut
  {NoStop}%
\bibitem [{\citenamefont {P.~Busch}(2009)}]{bookLR}%
  \BibitemOpen
  \bibfield  {author} {\bibinfo {author} {\bibfnamefont {P.~L.}\ \bibnamefont
  {P.~Busch}},\ }\href@noop {} {\emph {\bibinfo {title} {L\"{u}ders Rule. In:
  D. Greenberger, K. Hentschel K, F. Weinert (eds) Compendium of Quantum
  Physics.}}}\ (\bibinfo  {publisher} {Springer, Berlin, Heidelberg},\ \bibinfo
  {year} {2009})\BibitemShut {NoStop}%
\bibitem [{\citenamefont {Nielsen}\ and\ \citenamefont
  {Chuang}(2010)}]{nielsen_chuang_2010}%
  \BibitemOpen
  \bibfield  {author} {\bibinfo {author} {\bibfnamefont {M.~A.}\ \bibnamefont
  {Nielsen}}\ and\ \bibinfo {author} {\bibfnamefont {I.~L.}\ \bibnamefont
  {Chuang}},\ }\href@noop {} {\emph {\bibinfo {title} {Quantum computation and
  quantum information}}}\ (\bibinfo  {publisher} {Cambridge University Press},\
  \bibinfo {year} {2010})\BibitemShut {NoStop}%
\bibitem [{\citenamefont
  {Lüders}(2006)}]{https://doi.org/10.1002/andp.200610207}%
  \BibitemOpen
  \bibfield  {author} {\bibinfo {author} {\bibfnamefont {G.}~\bibnamefont
  {Lüders}},\ }\href {\doibase https://doi.org/10.1002/andp.200610207}
  {\bibfield  {journal} {\bibinfo  {journal} {Annalen der Physik}\ }\textbf
  {\bibinfo {volume} {15}},\ \bibinfo {pages} {663} (\bibinfo {year}
  {2006})}\BibitemShut {NoStop}%
\bibitem [{\citenamefont {Aharonov}\ \emph {et~al.}(1964)\citenamefont
  {Aharonov}, \citenamefont {Bergmann},\ and\ \citenamefont
  {Lebowitz}}]{PhysRev.134.B1410}%
  \BibitemOpen
  \bibfield  {author} {\bibinfo {author} {\bibfnamefont {Y.}~\bibnamefont
  {Aharonov}}, \bibinfo {author} {\bibfnamefont {P.~G.}\ \bibnamefont
  {Bergmann}}, \ and\ \bibinfo {author} {\bibfnamefont {J.~L.}\ \bibnamefont
  {Lebowitz}},\ }\href {\doibase 10.1103/PhysRev.134.B1410} {\bibfield
  {journal} {\bibinfo  {journal} {Phys. Rev.}\ }\textbf {\bibinfo {volume}
  {134}},\ \bibinfo {pages} {B1410} (\bibinfo {year} {1964})}\BibitemShut
  {NoStop}%
\bibitem [{\citenamefont {Vaidman}\ \emph {et~al.}(2017)\citenamefont
  {Vaidman}, \citenamefont {Ben-Israel}, \citenamefont {Dziewior},
  \citenamefont {Knips}, \citenamefont {Wei\ss{}l}, \citenamefont {Meinecke},
  \citenamefont {Schwemmer}, \citenamefont {Ber},\ and\ \citenamefont
  {Weinfurter}}]{PhysRevA.96.032114}%
  \BibitemOpen
  \bibfield  {author} {\bibinfo {author} {\bibfnamefont {L.}~\bibnamefont
  {Vaidman}}, \bibinfo {author} {\bibfnamefont {A.}~\bibnamefont {Ben-Israel}},
  \bibinfo {author} {\bibfnamefont {J.}~\bibnamefont {Dziewior}}, \bibinfo
  {author} {\bibfnamefont {L.}~\bibnamefont {Knips}}, \bibinfo {author}
  {\bibfnamefont {M.}~\bibnamefont {Wei\ss{}l}}, \bibinfo {author}
  {\bibfnamefont {J.}~\bibnamefont {Meinecke}}, \bibinfo {author}
  {\bibfnamefont {C.}~\bibnamefont {Schwemmer}}, \bibinfo {author}
  {\bibfnamefont {R.}~\bibnamefont {Ber}}, \ and\ \bibinfo {author}
  {\bibfnamefont {H.}~\bibnamefont {Weinfurter}},\ }\href {\doibase
  10.1103/PhysRevA.96.032114} {\bibfield  {journal} {\bibinfo  {journal} {Phys.
  Rev. A}\ }\textbf {\bibinfo {volume} {96}},\ \bibinfo {pages} {032114}
  (\bibinfo {year} {2017})}\BibitemShut {NoStop}%
\bibitem [{\citenamefont {Ozawa}(2003)}]{PhysRevA.67.042105}%
  \BibitemOpen
  \bibfield  {author} {\bibinfo {author} {\bibfnamefont {M.}~\bibnamefont
  {Ozawa}},\ }\href {\doibase 10.1103/PhysRevA.67.042105} {\bibfield  {journal}
  {\bibinfo  {journal} {Phys. Rev. A}\ }\textbf {\bibinfo {volume} {67}},\
  \bibinfo {pages} {042105} (\bibinfo {year} {2003})}\BibitemShut {NoStop}%
\bibitem [{\citenamefont {Ozawa}(2004)}]{OZAWA2004350}%
  \BibitemOpen
  \bibfield  {author} {\bibinfo {author} {\bibfnamefont {M.}~\bibnamefont
  {Ozawa}},\ }\href {\doibase https://doi.org/10.1016/j.aop.2003.12.012}
  {\bibfield  {journal} {\bibinfo  {journal} {Annals of Physics}\ }\textbf
  {\bibinfo {volume} {311}},\ \bibinfo {pages} {350 } (\bibinfo {year}
  {2004})}\BibitemShut {NoStop}%
\bibitem [{\citenamefont {von Neumann}(1955)}]{neumann}%
  \BibitemOpen
  \bibfield  {author} {\bibinfo {author} {\bibfnamefont {J.}~\bibnamefont {von
  Neumann}},\ }\href@noop {} {\emph {\bibinfo {title} {Mathematische Grundlagen
  der Quantenmechanik}}}\ (\bibinfo  {publisher} {Springer, Berlin, 1932;
  English translation Princeton University Press, Princeton, NJ, 1955},\
  \bibinfo {year} {1955})\BibitemShut {NoStop}%
\bibitem [{Note1()}]{Note1}%
  \BibitemOpen
  \bibinfo {note} {Using the Baker-Campbell-Hausdorff formula \cite {Hall2015},
  we expand $ \protect \mathbfcal U^\dagger \protect \mathbfcal O_0\ \protect
  \mathbfcal U = \protect \mathbfcal O_0 + it \leavevmode@ifvmode {\setbox \z@
  \hbox {\mathsurround \z@ $\nulldelimiterspace \z@ \left [\vcenter to\@ne
  \big@size {}\right .$}\box \z@ }\protect \bm {H}_\protect \mathcal {S}\otimes
  \protect \bm {H}_\protect \mathcal {M}, \protect \mathbfcal
  O_0\leavevmode@ifvmode {\setbox \z@ \hbox {\mathsurround \z@
  $\nulldelimiterspace \z@ \left ]\vcenter to\@ne \big@size {}\right .$}\box
  \z@ } + \protect \frac {(it)^2}{2!} \leavevmode@ifvmode {\setbox \z@ \hbox
  {\mathsurround \z@ $\nulldelimiterspace \z@ \left [\vcenter to1.5\big@size
  {}\right .$}\box \z@ }\protect \bm {H}_\protect \mathcal {S}\otimes \protect
  \bm {H}_\protect \mathcal {M}, \leavevmode@ifvmode {\setbox \z@ \hbox
  {\mathsurround \z@ $\nulldelimiterspace \z@ \left [\vcenter to\@ne \big@size
  {}\right .$}\box \z@ }\protect \bm {H}_\protect \mathcal {S}\otimes \protect
  \bm {H}_\protect \mathcal {M}, \protect \mathbfcal O_0\leavevmode@ifvmode
  {\setbox \z@ \hbox {\mathsurround \z@ $\nulldelimiterspace \z@ \left
  ]\vcenter to\@ne \big@size {}\right .$}\box \z@ } \leavevmode@ifvmode
  {\setbox \z@ \hbox {\mathsurround \z@ $\nulldelimiterspace \z@ \left
  ]\vcenter to1.5\big@size {}\right .$}\box \z@ } + \protect \cdots $. For
  small $t$ (or weak interaction), it yields $ \protect \mathbfcal U^\dagger
  \protect \mathbfcal O_0\ \protect \mathbfcal U \approx \protect \mathbfcal
  O_0 + it \leavevmode@ifvmode {\setbox \z@ \hbox {\mathsurround \z@
  $\nulldelimiterspace \z@ \left [\vcenter to\@ne \big@size {}\right .$}\box
  \z@ }\protect \bm {H}_\protect \mathcal {S} \otimes \protect \bm {H}_\protect
  \mathcal {M}, \protect \mathbfcal O_0\leavevmode@ifvmode {\setbox \z@ \hbox
  {\mathsurround \z@ $\nulldelimiterspace \z@ \left ]\vcenter to\@ne \big@size
  {}\right .$}\box \z@ } = \DOTSB \sum@ \slimits@ _k \protect \bm {S}_k\otimes
  \protect \bm {M}_k$.}\BibitemShut {Stop}%
\bibitem [{\citenamefont {Ozawa}(2005)}]{Ozawa_2005}%
  \BibitemOpen
  \bibfield  {author} {\bibinfo {author} {\bibfnamefont {M.}~\bibnamefont
  {Ozawa}},\ }\href {\doibase 10.1088/1464-4266/7/12/033} {\bibfield  {journal}
  {\bibinfo  {journal} {Journal of Optics B: Quantum and Semiclassical Optics}\
  }\textbf {\bibinfo {volume} {7}},\ \bibinfo {pages} {S672} (\bibinfo {year}
  {2005})}\BibitemShut {NoStop}%
\bibitem [{\citenamefont {Hall}(2015)}]{Hall2015}%
  \BibitemOpen
  \bibfield  {author} {\bibinfo {author} {\bibfnamefont {B.}~\bibnamefont
  {Hall}},\ }\enquote {\bibinfo {title} {The baker--campbell--hausdorff formula
  and its consequences},}\ in\ \href {\doibase 10.1007/978-3-319-13467-3_5}
  {\emph {\bibinfo {booktitle} {Lie Groups, Lie Algebras, and Representations:
  An Elementary Introduction}}}\ (\bibinfo  {publisher} {Springer International
  Publishing},\ \bibinfo {address} {Cham},\ \bibinfo {year} {2015})\ pp.\
  \bibinfo {pages} {109--137}\BibitemShut {NoStop}%
\end{thebibliography}%
\end{document}